\documentclass[aps,prl,twocolumn,superscriptaddress,]{revtex4-1}
\usepackage{amsmath}
\usepackage{amsfonts}
\usepackage{graphicx}
\usepackage{color}
\usepackage{dsfont}
\usepackage{verbatim}
\usepackage{mathtools}
\usepackage[normalem]{ulem}
\usepackage{comment}
\usepackage{stackrel}
\usepackage{bm}
\usepackage[pdftex]{hyperref}
\hypersetup{colorlinks = true, urlcolor = black, linkcolor = black, citecolor = black}

\definecolor{myblue}{rgb}{.93, .93, 1}

\newcommand{\bsub}{\begin{subequations}}
	\newcommand{\esub}{\end{subequations}}

\newcommand{\vex}[1]{\bm{\mathrm{#1}}}

\begin{document}

	\title{Enhanced superconductivity through virtual tunneling in Bernal bilayer graphene coupled to WSe$_2$}
	
	\author{Yang-Zhi~Chou}\email{yzchou@umd.edu}
	\affiliation{Condensed Matter Theory Center and Joint Quantum Institute, Department of Physics, University of Maryland, College Park, Maryland 20742, USA}

	\author{Fengcheng~Wu}
	\affiliation{School of Physics and Technology, Wuhan University, Wuhan  430072, China}
	\affiliation{Wuhan Institute of Quantum Technology, Wuhan 430206, China}
	
	\author{Sankar Das~Sarma}
	\affiliation{Condensed Matter Theory Center and Joint Quantum Institute, Department of Physics, University of Maryland, College Park, Maryland 20742, USA}	
	\date{\today}
	
	\begin{abstract}
		Motivated by a recent experiment [arXiv:2205.05087], we investigate a possible mechanism that enhances superconductivity in hole-doped Bernal bilayer graphene due to a proximate WSe$_2$ monolayer. We show that the virtual tunneling between WSe$_2$ and Bernal bilayer graphene, which is known to induce Ising spin-orbit coupling, can generate an additional attraction between two holes, providing a potential explanation for enhancing superconductivity in Bernal bilayer graphene. Using microscopic interlayer tunneling, we derive the Ising spin-orbit coupling and the effective attraction as functions of the twist angle between Bernal bilayer graphene and the WSe$_2$ monolayer. Our theory provides an intuitive and physical explanation for the intertwined relation between Ising spin-orbit coupling and superconductivity enhancement, which should motivate future studies.
	\end{abstract}
	
	\maketitle
	
	\textit{Introduction.---} Recent experiments on Bernal bilayer graphene (BBG) \cite{Zhou2022_BBG} and rhombohedral trilayer graphene (RTG) \cite{Zhou2021_SC_RTG} reveal multiple symmetry broken phases and provide a new understanding for superconductivity in general graphene systems (i.e., moir\'e \cite{Cao2018_tbg2,Yankowitz2019,Lu2019,Hao2021electric,Park2021tunable,Cao2021,Liu2021coulomb,Zhang2021ascendance,Park2021MAMG,Burg2022emergence,Siriviboon2021} or moir\'eless systems \cite{Zhou2021_SC_RTG,Zhou2022_BBG,Zhang2022SOCBBGSC}). In these non-moir\'e crystalline graphene multilayers, superconducting states with $T_c\le 0.1$K are found in narrow regions close to interaction-driven ``isospin'' polarized phases \cite{Zhou2021,Zhou2021_SC_RTG,Zhou2022_BBG}. 
	Theoretically, the acoustic-phonon-mediated pairing can provide a consistent resolution for superconductivity in BBG and RTG \cite{Chou2021_RTG_SC,Chou2022_BBG,Chou2022Eliashberg}, while interaction-driven mechanisms are also proposed \cite{Chatterjee2021,Ghazaryan2021,Dong2021,Cea2022,Szabo2022,You2022,Qin2022,Dai2022,Szabo2022BBG,Dong2022}.
	
	A new experiment on superconductivity in BBG demonstrates that superconductivity can be significantly enhanced with a proximate WSe$_2$ \cite{Zhang2022SOCBBGSC}. The system consists of a WSe$_2$ monolayer on top of a BBG, and a displacement field ($D$) is used to control the layer-polarization of the low-energy bands.
	For a sufficiently large $D>0$, the carriers of hole-doped BBG reside entirely on the top layer, and superconductivity is observed around $0.3$K without a magnetic field. The superconducting state shows a nontrivial response to an in-plane magnetic field -- a Pauli-limit violation at lower doping and Pauli-limited behavior at higher doping. For $D<0$, no superconductivity is observed for $T\ge 30$mK, but the normal states are essentially consistent with the previous experiment without a WSe$_2$ layer \cite{Zhou2022_BBG}. The strikingly different results for $D>0$ and $D<0$ suggest the significance of the proximate WSe$_2$ layer. It is important to emphasize that WSe$_2$ enhances superconductivity quite substantially -- the superconducting temperature is enhanced by an order of magnitude (from 30mK to 300mK), the region of the superconducting phase also becomes wider, and an in-plane magnetic field is no longer required to induce superconductivity. 
	
	\begin{figure}[t!]
		\includegraphics[width=0.45\textwidth]{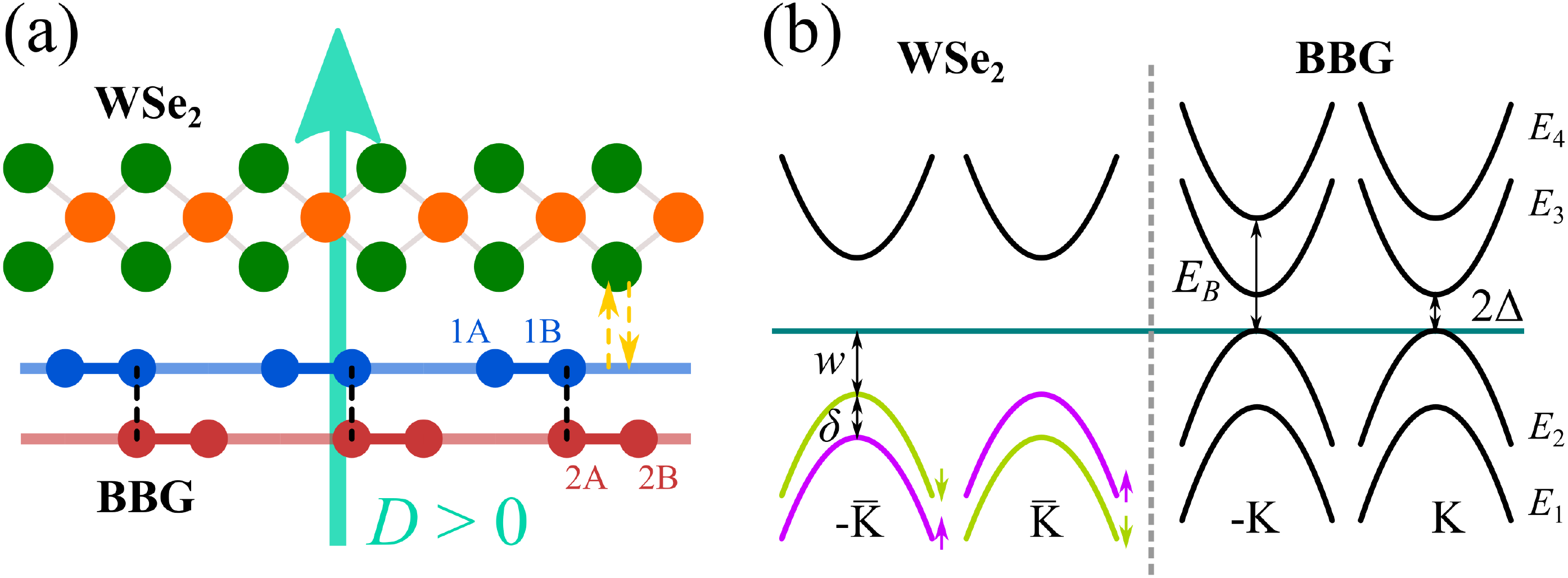}
		\caption{Setup of BBG-WSe$_2$ and band structure. (a) Side view of the BBG-WSe$_2$ system. The WSe$_2$ monolayer is on top of the BBG. A displacement field along the $z$-direction is exerted. (b) The schematic band structures of WSe$_2$ and BBG. The green line indicates the Fermi energy, which is on the band edge of the first BBG valence band ($E_2$). We ignore the spin splitting of the WSe$_2$ conduction bands in this illustration. We use $E_1$, $E_2$, $E_3$, and $E_4$ to label the BBG bands in ascending order in energy.}
		\label{Fig:Setup}
	\end{figure}
	
	The key task is to identify the physical origin of superconductivity enhancement.
	Since a small observable $T_c$ has been found in BBG \cite{Zhou2022_BBG}, any additional pairing glue or reduction of Coulomb repulsion can result in a noticeable enhancement in superconductivity. However, such a cooperative enhancing mechanism must be absent without a nearby WSe$_2$ layer, manifesting an asymmetric effect in $D>0$ and $D<0$. The main goal of the current work is to provide a potential explanation for the WSe$_2$-enhanced superconductivity in BBG \cite{Zhang2022SOCBBGSC}.
	
	In this Letter, we propose a \textit{novel} mechanism that enhances pairings in a BBG-WSe$_2$ system based on the interlayer tunneling between WSe$_2$ and BBG. Such a tunneling process is believed to induce Ising spin-orbit coupling (ISOC) in BBG \cite{Island2019spin,Zhang2022SOCBBGSC,Gmitra2016,Gmitra2017,Khoo2017demand}, implying significant interlayer tunneling. We develop a minimal theory that produces an effective attraction between two holes in the slightly hole-doped BBG via a virtual interlayer tunneling process in combination with an interaction between hole carriers and the virtual electron. Furthermore, we derive the ISOC and effective attraction as functions of the relative angle between WSe$_2$ and BBG, incorporating microscopic tunneling at extended Brillouin zones \cite{Bistritzer2010,David2019}. Our results suggest that the enhanced superconductivity can be explained by the virtual tunneling from the WSe$_2$ layer in cooperation with the electron-phonon interaction, paving the way for higher-$T_c$ superconducting states in graphene systems.

	\textit{Model.---} We are interested in a BBG-WSe$_2$ system as depicted in Fig.~\ref{Fig:Setup}. In the presence of a sufficiently large displacement field along $z$-direction ($D>0$), the low-energy valence band of BBG is polarized at the 1A site [illustrated in Fig.~\ref{Fig:Setup}(a)] on the top graphene layer. It was shown theoretically \cite{Gmitra2017,Khoo2017demand} and experimentally \cite{Island2019spin,Zhang2022SOCBBGSC} that ISOC is induced primarily on the layer proximate to WSe$_2$, suggesting that tunneling between WSe$_2$ and the top graphene layer is essential. Thus, a minimal model must include certain properties of WSe$_2$ and BBG bands as well as the interlayer tunneling between WSe$_2$ and the top layer of BBG.

	\begin{figure}[t!]
		\includegraphics[width=0.47\textwidth]{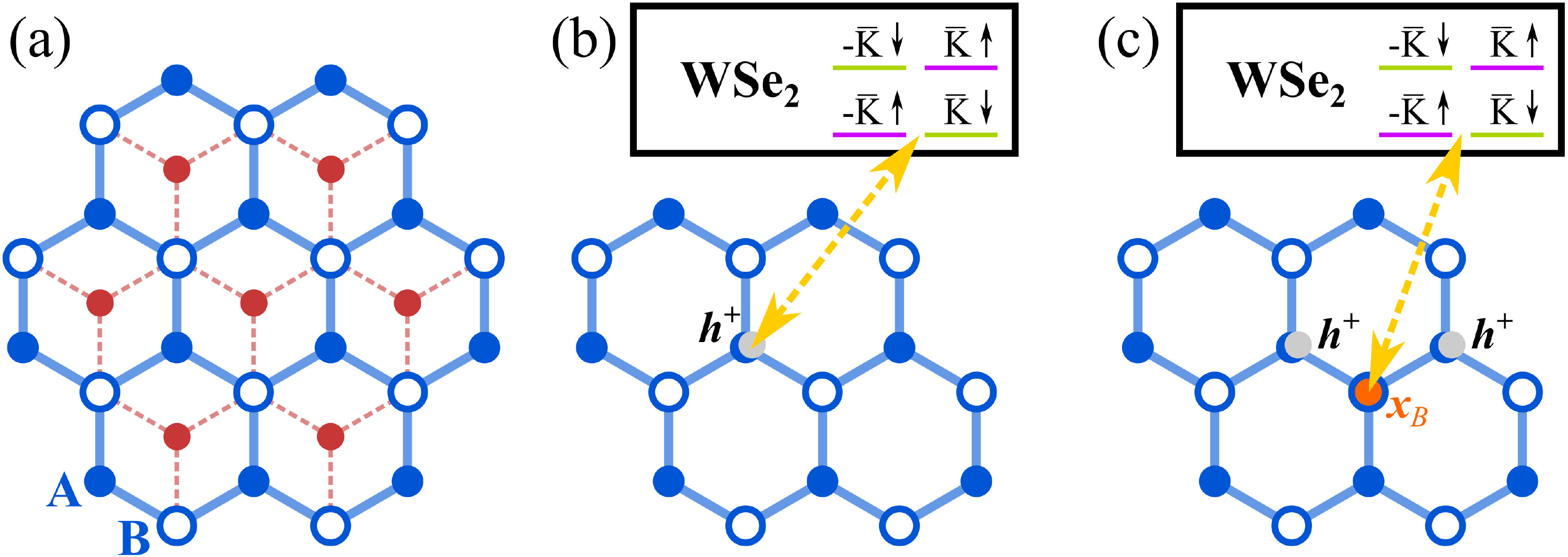}
		\caption{Lattice model and virtual tunneling processes. (a) The effective honeycomb lattice model for BBG. A sites (blue dots) and B sites (blue opened circles) correspond to the positions of 1A and 1B sites in Fig.~\ref{Fig:Setup}(a), respectively; the red dots indicate the 2B sites in the bottom graphene layer; 2A sites are right below the 1B sites. (b) ISOC due to virtual tunneling. 
			(c) Attraction due to virtual tunneling. }
		\label{Fig:Lattice}
	\end{figure}	
	
	To simplify the problem, we consider an \textit{effective} honeycomb model as follows (see Fig.~\ref{Fig:Lattice} and \cite{SM}):
	\begin{align}\label{Eq:H_tG}
		\hat{H}_{\text{tG}}=E_A\sum_{\vex{r}_A}n(\vex{r}_A)+E_B\sum_{\vex{r}_B}n(\vex{r}_B),
	\end{align}
	where $E_A$ ($E_B$) corresponds to the onsite energy of the effective A (B) sites, $n(\vex{r}_{\sigma})=\sum_{\tau,s}c^{\dagger}_{\tau \sigma s}(\vex{r}_{\sigma})c_{\tau \sigma s}(\vex{r}_{\sigma})$ is the number operator at site $\vex{r}_{\sigma}$, $c_{\tau \sigma s}$ is the fermionic annihilation operator with valley $\tau K$, sublattice $\sigma=A,B$, and spin $s$. The lack of hopping is because we consider momentum right at $K$ or $-K$, where the system can be viewed as a collection of decoupled atomic sites. Equation~(\ref{Eq:H_tG}) is a simplified description of BBG degrees of freedom relevant to the virtual tunneling processes considered in this work, and we retain only the $E_2$ band (A sites) and the $E_4$ band (B sites), where the $E_2$ and $E_4$ bands are labeled in Fig.~\ref{Fig:Setup}(b). Due to the interlayer dimerization between 1B and 2A sites, the microscopic 1B sites of BBG are associated with both the $E_1$ and $E_4$ bands. For our proposal, the $E_4$ band is important, while the $E_1$ band is ignored.
	Thus, we consider $E_A=0$ and set the value of $E_B$ to the energy difference between the $E_4$ and the $E_2$ band edges \cite{EB}, as illustrated in Fig.~\ref{Fig:Setup}(b).
	In such a model, the charge neutral configuration [i.e., $\mathcal{E}_F$ is inside the $2\Delta$ gap of Fig.~\ref{Fig:Setup}(b)] corresponds to a state with completely filled A sites and empty B sites. In our case with $\mathcal{E}_F$ at the $E_2$ band edge, the system is slightly hole-doped, and we can consider ground states with dilute holes on the A sites of the effective honeycomb lattice model [given by Eq.~(\ref{Eq:H_tG})]. Again, the effective description here is valid when $\mathcal{E}_F$ is at the $E_2$ band edge.

	In addition to the onsite potential, we consider electron-electron interactions given by
	\begin{align}\label{Eq:H_I}
		\hat{H}_I\!=\!\frac{U_0}{2}\sum_{\vex{r}} \delta n(\vex{r})\left[\delta n(\vex{r})\!-\!1\right]\!+\!U_1\!\!\!\!\sum_{\langle \vex{r}_A,\vex{r}_B\rangle}\!\!\!\!\delta n(\vex{r}_A)\delta n(\vex{r}_B),
	\end{align}
	where $\delta n(\vex{r})=n(\vex{r})-\langle 0| n(\vex{r})|0\rangle$, $|0\rangle$ is the state with a charge neutral configuration (i.e., filled $A$ sites and empty $B$ sites), $U_0>0$ ($U_1>0$) is the onsite (nearest-neighbor) Coulomb interaction, and $\langle \vex{r}_A,\vex{r}_B\rangle$ denotes the nearest-neighbor pair. We consider a sufficiently large $U_0$ such that at most one hole (electrons) can be created on sublattice A (B). The $U_1$ term describes the interaction between nearest-neighbor sites, and $U_1<E_B$ is assumed (as the spontaneous formation of dipoles is forbidden). We will focus on the electron-hole attraction betweenan electron on the B site and a hole on the nearest-neighbor A site in the virtual process.
	
	The WSe$_2$ layer can be described by a semiconductor bandstructure with spin split valence bands \cite{Xiao2012} as illustrated in Fig.~\ref{Fig:Setup}(b). Specifically, the energy splittings can be described by an ISOC, $\lambda\tau_z s_z$, with $\tau_z$ ($s_z$) being the $z$-component Pauli matrix for valley (spin). The interlayer tunneling between WSe$_2$ and BBG can facilitate spin-orbit splitting in BBG valence bands. To provide an intuitive understanding, we treat WSe$_2$ valence bands as a few representative energy levels described by a simplified Hamiltonian,
	\begin{align}\label{Eq:H_d}
		\hat{H}_{d}\!=\!-\!W\!\!\left(d_{+,\uparrow}^{\dagger}d_{+,\uparrow}\!+\!d_{-,\downarrow}^{\dagger}d_{-,\downarrow}\right)\!-\!(W+\delta)\!\!\left(d_{-,\uparrow}^{\dagger}d_{-,\uparrow}\!+\!d_{+,\downarrow}^{\dagger}d_{+,\downarrow}\right)\!,
	\end{align}
	where $d_{\tau,s}$ denotes the fermionic annihilation operator with valley $\tau \bar{K}$ and spin $s$ in the WSe$_2$ valence bands. $W$ and $\delta$ are the parameters for the WSe$_2$ valence bands. 
	
	Finally, we consider a tunneling Hamiltonian between WSe$_2$ and BBG given by
	\begin{align}\label{Eq:H_V}
		\nonumber\hat{H}_V\!=\!\!&\sum_{\tau,s,\vex{r}_A}\!\!\left\{c^{\dagger}_{\tau A s}(\vex{r}_A)\left[V^{A}_{\tau s}d_{\tau, s}+\bar{V}^{A}_{\tau s}d_{-\tau, s}\right]\!+\text{H.c.}\right\}\\
		&+\!\!\sum_{\tau,s,\vex{r}_B}\!\!\left\{c^{\dagger}_{\tau B s}(\vex{r}_B)\left[V^{B}_{\tau s}d_{\tau, s}\!+\!\bar{V}^{B}_{\tau s}d_{-\tau, s}\right]\!+\text{H.c.}\right\}\!,
	\end{align}
	where $V^{\sigma}_{\tau s}$ is the tunneling strength for the intravalley process (i.e., $\tau \bar{K}$ to $\tau K$), and $\bar{V}^{\sigma}_{\tau s}$ is the tunneling strength for the intervalley process (i.e., $-\tau \bar{K}$ to $\tau K$) \cite{Tunneling_V}. 
	Since the entire system preserves the (spinful) time-reversal symmetry, the tunneling terms obey $\left|V^{\sigma}_{\bar{\tau} \bar{s}}\right|=\left|V^{\sigma}_{\tau s}\right|$ and $\left|\bar{V}^{\sigma}_{\bar{\tau} \bar{s}}\right|=\left|\bar{V}^{\sigma}_{\tau s}\right|$, where $\bar{\tau}$ ($\bar{s}$) means the time-reversal partner of $\tau$ ($s$).
	
	The model designed here is physically motivated for understanding ISOC and effective attraction via interlayer tunneling. However, such a simplified model cannot capture the full microscopic detail. We will later use a detailed approach incorporating WSe$_2$ band structures and the microscopic interlayer tunneling matrix elements. 
	
	\begin{figure}[t!]
		\includegraphics[width=0.45\textwidth]{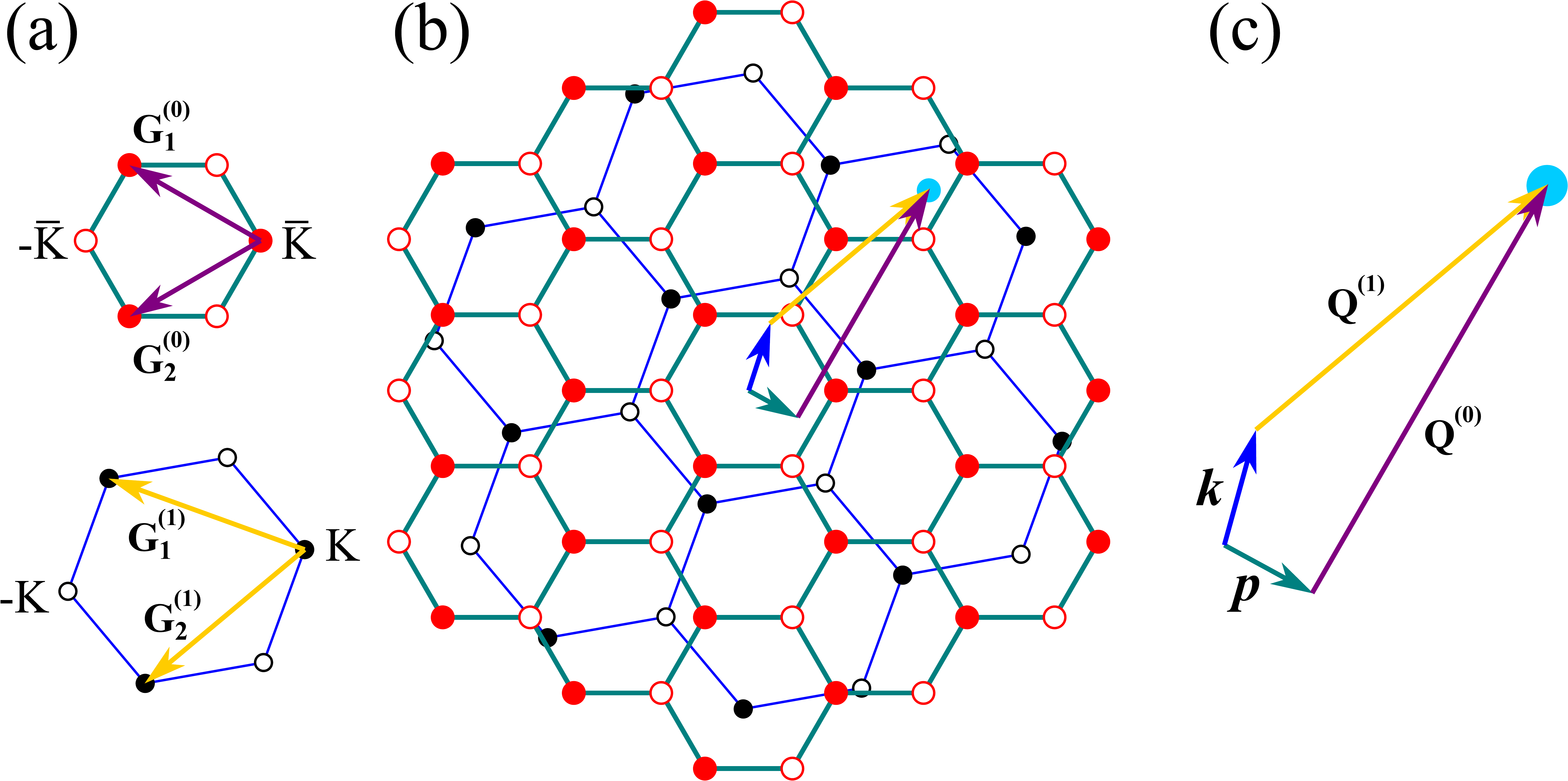}
		\caption{Brillouin zone and geometry of interlayer tunneling. (a) First Brillouin zone of WSe$_2$ (green) and BBG (blue). BBG is rotated by a twist angle $\theta$. We use $\theta=10^{\circ}$ here. $\vex{G}^{(0)}_a$'s and $\vex{G}^{(1)}_a$'s are the primitive lattice vectors of a reciprocal lattice of WSe$_2$ and BBG, respectively. (b) Extended Brillouin zones. (c) Momenta $\vex{p}$ and $\vex{k}$ have the same crystal momentum as $\vex{p}+\vex{Q}^{(0)}=\vex{k}+\vex{Q}^{(1)}$, where $\vex{Q}^{(0)}$ and $\vex{Q}^{(1)}$ are the reciprocal lattice vectors of WSe$_2$ and BBG, respectively.}
		\label{Fig:BZ_main}
	\end{figure}

	\textit{Ising spin-orbit coupling and effective attraction.---} The minimal model $\hat{H}=\hat{H}_{\text{tG}}+\hat{H}_{I}+\hat{H}_d+\hat{H}_V$ [given by Eqs.~(\ref{Eq:H_tG}), (\ref{Eq:H_I}), (\ref{Eq:H_d}), and (\ref{Eq:H_V})] can straightforwardly produce ISOC in the first valence band of BBG. The main idea is that the second-order perturbation in $\hat{H}_V$ as sketched in Fig.~\ref{Fig:Lattice}(b) (see also \cite{SM}) generates spin-valley-splitting energy levels on the A sites, realizing an ISOC with strength
	\begin{align}\label{Eq:lambda_I}
		\lambda_I=&\frac{\left|V^A_{+\uparrow}\right|^2-\left|\bar{V}^A_{+\uparrow}\right|^2}{W}
		-\frac{\left|V^A_{+\downarrow}\right|^2-\left|\bar{V}^A_{+\downarrow}\right|^2}{W+\delta}.
	\end{align} 
	
	The main goal of this work is to investigate if the interlayer tunneling between WSe$_2$ and BBG can generate an effective attractive interaction. Specifically, we consider two holes that contain a common nearest-neighbor B site at $\vex{x}_B$ as illustrated in Fig.~\ref{Fig:Lattice}(c). At the second-order perturbation of $\hat{H}_V$, the interplay between virtual tunneling and the interaction $\hat{H}_I$ [Eq.~(\ref{Eq:H_I})] generates an effective attraction between the holes, described by \cite{SM}
	\begin{align}\label{Eq:H_att}
		\hat{H}_{\text{att}}=-\mathcal{U}_{\text{eff}}\sum_{\langle\langle \vex{r}_A,\vex{r}_A'\rangle\rangle}\delta n(\vex{r}_A)\delta n(\vex{r}_A'),
	\end{align}
	where the sum runs over the nearby pairs on A sites and
	\begin{align}\label{Eq:U}
		\nonumber\mathcal{U}_{\text{eff}}\!=&\frac{\mathcal{V}^2_{\uparrow}}{W+E_B}+\frac{\mathcal{V}^2_{\uparrow}}{W+E_B-2U_1}-\frac{2\mathcal{V}^2_{\uparrow}}{W+E_B-U_1}\\
		&\!+\!\frac{\mathcal{V}^2_{\downarrow}}{W+\delta+E_B}\!+\!\frac{\mathcal{V}^2_{\downarrow}}{W+\delta+E_B-2U_1}\!-\!\frac{2\mathcal{V}^2_{\downarrow}}{W+\delta+E_B-U_1}
	\end{align}
	with $\mathcal{V}^2_{s}=2\left|V^{B}_{+s}\right|^2+2\left|\bar{V}^{B}_{+s}\right|^2$. $\mathcal{U}_{\text{eff}}$ vanishes as $U_1$ is absent.
	While the electron on a B site has a large local energy $E_B$, the nearest-neighbor electron-hole attraction can lower the total energy in a virtual state. In Eq.~(\ref{Eq:U}), those terms with $-2U_1$, corresponding to virtual tunneling to $\vex{x}_B$, yield the dominant contributions as long as $U_1>0$. We assume that $W+E_B>2U_1$ so that the any charge transfer between WSe$_2$ and BBG should be absent. Since $|V^A_{\tau s}|$ and $|V^B_{\tau s}|$ are of the same order of magnitude \cite{SM}, we expect that the virtual tunneling generates a sizable effective attraction $\mathcal{U}_{\text{eff}}$. 
	Our theory therefore provides an intuitive understanding of the effective attraction due to a proximate WSe$_2$ layer.

	The proposed mechanism here is conceptually related to the ``polarizer'' idea \cite{Little1964,Hamo2016electron} and the repulsion-induced attraction in models on the honeycomb lattice \cite{Slagle2020,Crepel2021,Crepel2022PNAS,Crepel2022}. We discuss a few differences between our work and Refs. \cite{Slagle2020,Crepel2021,Crepel2022PNAS,Crepel2022} -- (i) The virtual process is due to interlayer tunneling rather than intralayer hopping, and (ii) the electron-hole attraction in the virtual process rather than electron-electron repulsion. Point (i) is crucial as our mechanism describes a possible enhanced attraction from WSe$_2$ rather than pairings due to intralayer processes. In addition, point (ii) allows for a wider parameter range for a sizable effective attraction because the large onsite energy $E_B$ can be compensated by a nearest neighbor attraction.

	\textit{Interlayer tunneling and twist angle.---} The interlayer tunneling between WSe$_2$ and BBG crucially determines ISOC as well as the virtual-tunneling-induced effective attraction. The interlayer tunneling preserves crystal momentum as the matrix element primarily depends on the distance between the sites. Within the two-center approximation scheme, the tunneling between the layers can be described by \cite{Bistritzer2010}
	\begin{align}
		\nonumber T_{\vex{k},\vex{p}}^{\alpha,\sigma;\beta,\sigma'}=&\frac{1}{\sqrt{\mathcal{A}_0\mathcal{A}_1}}\left(a^{(1)}_{\vex{k},\alpha,\sigma}\right)^*\left(a^{(0)}_{\vex{p},\beta,\sigma'}\right)\times\\
		\label{Eq:T_inter}&\!\!\sum_{\vex{Q}^{(1)},\vex{Q}^{(0)}}\!\!\tilde{t}_{\vex{k}+\vex{Q}^{(1)}}e^{i\phi^{\sigma;\sigma'}_{\vex{Q}^{(1)},\vex{Q}^{(0)}}}\delta_{\vex{k}+\vex{Q}^{(1)},\vex{p}+\vex{Q}^{(0)}},
	\end{align}
	where $\mathcal{A}_0$ ($\mathcal{A}_1$) is the unit-cell area of WSe$_2$ (BBG), $\alpha$ and $\beta$ are the band indexes, $\vex{k}$ of BBG and $\vex{p}$ of WSe$_2$ are the momentum relative to $\Gamma$ point (Brillouin zone center), $\sigma$ and $\sigma'$ are the sublattice (orbital) indexes, $a^{(l)}_{\vex{k},\alpha,\sigma}$ is the sublattice (orbital) projection of a Bloch state $|\vex{k},\alpha\rangle$ at layer $l$, $\vex{Q}^{(0)}$, $\vex{Q}^{(1)}$ are the reciprocal lattice vectors in WSe$_2$ and BBG, respectively, and $\phi^{\sigma;\sigma'}_{\vex{Q}^{(1)},\vex{Q}^{(0)}}$ is a phase factor depending on sublattice. In the above expression, we use index $0$ for the WSe$_2$ layer and index $1$ for the top graphene layer of BBG. $\tilde{t}_{\vex{k}}$ is the 2D Fourier transform of the interlayer tunneling amplitude with a finite range, and we use a stretched exponential ansatz form \cite{Bistritzer2010},
	\begin{align}\label{Eq:t_k}
		|\tilde{t}_{\vex{k}}|=t_0 e^{-\chi\left(|\vex{k}|z_{\perp}\right)^{\gamma}},
	\end{align}
	where $t_0$ is an overall constant, $\chi$ is an order 1 numerical constant, $\gamma$ is the exponent of the stretched exponential, and $z_{\perp}$ is the distance between the WSe$_2$ monolayer and the top graphene layer of BBG \cite{zperp}. Note that Eq.~(\ref{Eq:t_k}) is an empirical expression for a finite $\vex{k}$, and the potential complications for $\vex{k}\rightarrow 0$ are not relevant to our problem.

	The interlayer tunneling described by Eq.~(\ref{Eq:T_inter}) is a highly nontrivial single-particle process. Microscopically, WSe$_2$ and BBG have a relative angle that is tunable experimentally as well as different lattice constants ($d=3.31\text{\AA}$ for WSe$_2$, $a=2.46\text{\AA}$ for BBG), resulting in Brillouin zones illustrated in Fig.~\ref{Fig:BZ_main}. Note that $T_{\vex{k},\vex{p}}^{\alpha,\sigma;\beta,\sigma'}$ is nonzero as long as $\vex{k}+\vex{Q}^{(1)}=\vex{p}+\vex{Q}^{(0)}$ for some reciprocal lattice vectors $\vex{Q}^{(0)}$ and $\vex{Q}^{(1)}$. Thus, a full calculation must incorporate a sufficiently large number of extended Brillouin zones.
	Using the virtual-tunneling ideas discussed previously, we can compute the $\lambda_I$ and $\mathcal{U}_{\text{eff}}$ incorporating the microscopic interlayer tunneling matrix elements between the $\vex{k}\cdot\vex{p}$ WSe$_2$ bands \cite{Xiao2012} and our honeycomb model. We summarize the main results, and the complete derivations are provided in \cite{SM}.

\begin{figure}[t!]
		\includegraphics[width=0.35\textwidth]{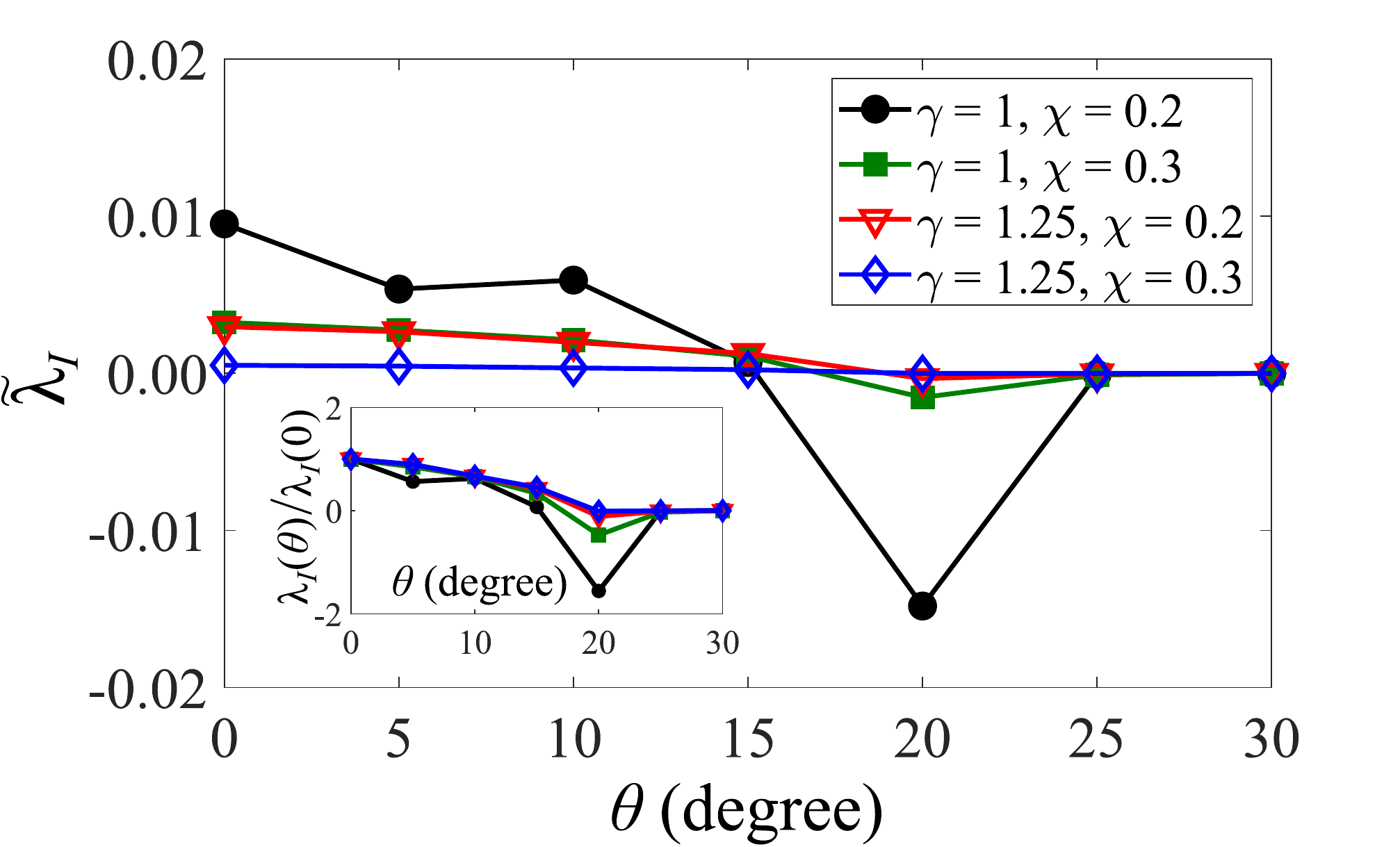}
		\caption{Ising spin-orbit coupling versus twist angle ($\theta$) with microscopic interlayer tunneling. We plot the dimensionless spin-orbit coupling $\tilde{\lambda}_I=\lambda_I \mathcal{A}_0\mathcal{A}_1 \text{eV}/t_0^2$ as a function of $\theta$ with different values of $\gamma$ and $\chi$. $W=0.62$eV and $\delta=0.46$eV for all the plots.
		}
		\label{Fig:SOC}
	\end{figure}

\begin{figure}[t!]
		\includegraphics[width=0.45\textwidth]{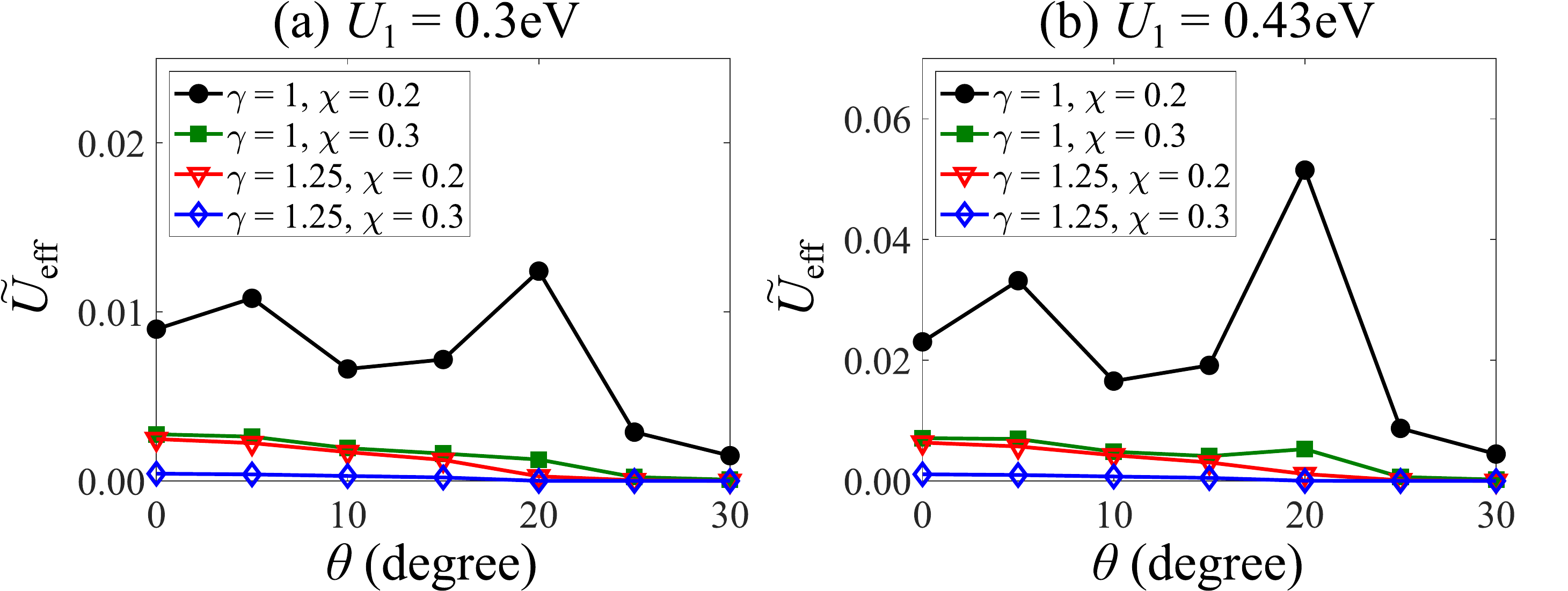}
		\caption{Effective attraction versus twist angle ($\theta$)  with microscopic interlayer tunneling. We plot the dimensionless spin-orbit coupling $\tilde{U}_{\text{eff}}=\mathcal{U}_{\text{eff}} \mathcal{A}_0\mathcal{A}_1 \text{eV}/t_0^2$ as a function of $\theta$ with different values of $\gamma$, $\chi$, and $U_1$. (a) $U_1=0.3$eV. (b) $U_1=0.43$eV. $W=0.62$eV and $\delta=0.46$eV for all the plots.
		}
		\label{Fig:attraction}
	\end{figure}

	In Fig.~\ref{Fig:SOC}, we plot the twist-angle ($\theta$) dependence of $\lambda_I$ with a few representative values of $\chi$ and $\gamma$ \cite{parameters}. $\lambda_I$ is generally positive for small positive $\theta$ and becomes negative slightly above $\theta=15^{\circ}$. 
	Moreover, 
	nonmonotonic behavior can manifest near $\theta=5^{\circ}$ and $\theta=20^{\circ}$ for smaller $\gamma$ and $\chi$. These results can be understood by the geometry of momentum at $\theta=5^{\circ}$ and $\theta\ge20^{\circ}$ \cite{SM}. 
	At $\theta=30^{\circ}$, $\lambda_I$ vanishes because the intervalley and intravalley tunnelings have exactly the same contributions. Similar nonmonotonic behavior near $\theta=20^{\circ}$ was also theoretically reported in graphene coupled to transition metal dichalcogenide \cite{David2019,Li2019_twist,Naimer2021}. In addition, the sign changing in $\lambda_I$ (curves with $\gamma=1$ in Fig.~\ref{Fig:SOC}) was also obtained in Ref.~\cite{David2019}.
	In Fig.~\ref{Fig:attraction}, we plot the $\theta$ dependence of $\mathcal{U}_{\text{eff}}$ with $U_1=0.3$eV and $U_1=0.43$eV (the largest possible value in our theory) and a few different values of $\chi$ and $\gamma$. $\mathcal{U}_{\text{eff}}$ is generally larger at smaller $\theta$, but nonmonotonic features might develop for small $\gamma$ and $\chi$. While the qualitative results are insensitive to $U_1$, the quantitative values depend on $U_1$ \cite{SM}. An important implication here is that ISOC strength and the effective attraction are \textit{not} directly related to each other. The full results depend a lot on the details of $T_{\vex{k},\vex{p}}^{\alpha,\sigma;\beta,\sigma'}$.
	
	\textit{Discussion.---} The virtual-tunneling-induced attraction might explain the enhanced superconductivity in BBG-WSe$_2$ \cite{Zhang2022SOCBBGSC} as it enters the theory nonperturbatively in an exponential function determining $T_c$ and the effect is primarily on $D>0$. In the BBG experiment without WSe$_2$ \cite{Zhou2022_BBG}, superconductivity with $T_c\sim 30$mK was reported at the doping density $n_e=-6\times 10^{12}$cm$^{-2}$ with a Fermi-surface degeneracy factor of 2, implying that the pairing interaction between holes overcomes Coulomb suppression. Since the enhanced superconductivity is found near the same doping density with a very similar Fermi-surface degeneracy factor, we anticipate that the same pairing mechanism manifests in BBG with or without the proximate WSe$_2$. According to Ref.~\cite{Chou2022_BBG}, the acoustic-phonon-mediated attraction is slightly stronger than the Coulomb suppression, so any additional pairing glue due to presence of the WSe$_2$, albeit small compared to the Coulomb repulsion, might considerably enhance superconductivity because $T_c$ is exponentially related to the coupling constant. As such, our theory based on the interplay between virtual tunneling and interaction is a possible explanation for the BBG-WSe$_2$ experiment \cite{Zhang2022SOCBBGSC}. We also derive the twist-angle dependence of the ISOC parameter and the effective attraction, providing guidance for future BBG-WSe$_2$ systems.

	It is worth mentioning that our theory most likely underestimates $\mathcal{U}_{\text{eff}}$ because we consider only the nearest-neighbor Coulomb interaction and only valley momenta (i.e., $K$ and $-K$ points) in the BBG bands. With a long-range Coulomb interaction, a long-range effective attraction can also arise from virtual tunneling processes, resulting in an overall stronger pairing. Moreover, incorporating all the allowed momenta in the BBG $E_4$ band also enhances $\mathcal{U}_{\text{eff}}$ because there are more states to be tunneled into.

	Now we discuss the pairing symmetry and the normal state in the BBG-WSe$_2$ system. We anticipate that intervalley intrasublattice pairings dominate \cite{Wu2019_phonon,Chou2021correlation,Chou2021_RTG_SC} because of the layer-sublattice polarization in BBG. Based on the previous work on acoustic phonons \cite{Chou2022_BBG} and the ideas discussed in this Letter, we assume that the dominant pairing is due to the acoustic phonons, and the virtual-tunneling-induced attraction is the subleading pairing glue. 
	Note that the normal states for superconductivity in Refs.~\cite{Zhou2022_BBG} and \cite{Zhang2022SOCBBGSC} are qualitatively different due to the differences in spin-orbit coupling and the in-plane magnetic field. While our theory does not determine the normal state properties, the major pairing glue, the acoustic-phonon-mediated pairing, has a $SU(2)\times SU(2)$ symmetry \cite{Wu2019_phonon,Chou2021_RTG_SC,Chou2022_BBG} (allowing for singlet, triplet, and singlet-triplet mixing). Thus, the superconductivity should be enhanced regardless of the details of the spin symmetry in the normal state or in the subleading pairing \cite{pairing_symmetry}.
	Due to the induced spin-orbit coupling, we anticipate an admixture of singlet and triplet pairings \cite{Lu2015evidence,Xi2016ising}, which can produce a beyond-Pauli-limit response to the in-plane magnetic field, consistent with the lower-doping superconductivity in the BBG-WSe$_2$ experiment \cite{Zhang2022SOCBBGSC}. 
	
	While our theory provides a potentially consistent resolution to the BBG-WSe$_2$ experiment, there are a few issues that require further investigations. First, the effective attraction $\mathcal{U}_{\text{eff}}$ depends on the value of $U_1$, which we treat as a parameter. This makes a quantitative estimate of $\mathcal{U}_{\text{eff}}$ difficult in our theory. The other issue is related to the differences in the normal states between Refs.~\cite{Zhou2022_BBG} and \cite{Zhang2022SOCBBGSC}. In our work, we simply assume that such differences are not crucial to the superconductivity enhancement. However, as pointed out in Ref.~\cite{Zhang2022SOCBBGSC}, a nontrivial normal state due to an interplay between the interaction and spin orbit coupling might explain the enhancement of superconductivity. To fully understand the factor-of-10 enhancement of $T_c$ in the BBG-WSe$_2$ experiment \cite{Zhang2022SOCBBGSC}, one needs to incorporate both the additional pairing glue and the change in normal states.

	Finally, we discuss implications for experiments. First, the twist-angle dependence of ISOC [Fig.~\ref{Fig:SOC}] and effective attraction [Fig.~\ref{Fig:attraction}] might provide insights on why enhanced superconductivity is not always achieved in BBG-WSe$_2$ experiments \cite{Zhang2022SOCBBGSC}. 
	Based on our theory, applying a pressure to the system should considerably raise $T_c$ since an applied pressure should enhance tunneling. Furthermore, one can use the ideas of this work to look for the optimal proximate layer for enhancing superconductivity. An important message of this Letter is that superconductivity can still be enhanced even if the induced spin-orbit coupling is very small, because of the exponential nature in $T_c$ and the complex relation between effective attraction and spin-orbit coupling.

	\begin{acknowledgments}
		\textit{Acknowledgments.---} We thank \'Etienne Lantagne-Hurtubise, Alex Thomson, and Cyprian Lewandowski for pointing out an issue in an earlier version of the manuscript. We are also grateful to Seth Davis, Jiabin Yu, and Ming Xie for useful discussions.
		This work is supported by the Laboratory for Physical Sciences (Y.-Z.C. and S.D.S.), by JQI-NSF-PFC (Y.-Z.C.), and by ARO W911NF2010232 (Y.-Z.C.). F.W. is supported by National Key R$\&$D Program of China 2021YFA1401300 and start-up funding of Wuhan University.
	\end{acknowledgments}


	\newpage \clearpage 
	
	\onecolumngrid
	
	\begin{center}
		{\large
			Enhanced superconductivity through virtual tunneling in Bernal bilayer graphene coupled to WSe$_2$
			\vspace{4pt}
			\\
			SUPPLEMENTAL MATERIAL
		}
	\end{center}
	
	\setcounter{figure}{0}
	\renewcommand{\thefigure}{S\arabic{figure}}
	\setcounter{equation}{0}
	\renewcommand{\theequation}{S\arabic{equation}}
	
	In this supplemental material, we provide some technical details for the main results in the main text.

	\section{Single-particle bands and effective honeycomb lattice model}
	
	The $\vex{k}\cdot\vex{p}$ Hamiltonian for BBG is based on Ref.~\cite{Jung2014S}. The $\hat{h}_{\tau}(\vex{k})$ in the main text is given by
	\begin{align}\label{Eq:h_k_BBG}
		\hat{h}_{\tau}(\vex{k})
		=\left[\begin{array}{cccc}
			-\Delta & v_0\Pi^{\dagger}(\vex{k}) & -v_4\Pi^{\dagger}(\vex{k}) & -v_3\Pi(\vex{k}) \\[2mm]
			v_0\Pi(\vex{k}) & \Delta'-\Delta & t_1 & -v_4\Pi^{\dagger}(\vex{k}) \\[2mm]
			-v_4\Pi(\vex{k})& t_1 & \Delta'+\Delta & v_0\Pi^{\dagger}(\vex{k}) \\[2mm]
			-v_3\Pi^{\dagger}(\vex{k})& -v_4\Pi(\vex{k}) & v_0\Pi(\vex{k})  & \Delta
		\end{array}
		\right],
	\end{align}
	where $\Pi(\vex{k})=\tau k_xa+ik_ya$, $a=2.46$\AA is the lattice constant of graphene and $\Delta$ encodes the electric potential difference from the displacement field. In this work, we take $\Delta=50$meV. Note that $\vex{k}$ is relative to $\tau K$. Other parameters are given by $v_0=2.261$ eV, $v_3=0.245$ eV, $v_4=0.12$ eV, $t_1=0.361$ eV, and $\Delta'=0.015$ eV. The basis of the matrix is (1A,1B,2A,2B).\\
	
	The low-energy bands manifest layer-sublattice polarization, so that only the A sites of the top layer (1A) and B sites of the bottom layer (2B) are essential for slightly doped BBG. This property arises from the interlayer nearest-neighbor tunnelings which tend to form dimerized bonds between 1B and 2A sites. For $\Delta>0$, the low-energy valence band can be well described by 1A sites alone. To see this, we examine $\vex{k}=0$ case in the following. Right at $\tau K$,
	\begin{align}
		\hat{h}_{\tau}(0)
		=\left[\begin{array}{cccc}
			-\Delta & 0 & 0 & 0 \\[2mm]
			0 & \Delta'-\Delta & t_1 & 0 \\[2mm]
			0 & t_1 & \Delta'+\Delta & 0 \\[2mm]
			0 & 0 & 0 & \Delta
		\end{array}
		\right].
	\end{align}
	The eigenvalues of $\hat{h}_{\tau}(0)$ are 
	\begin{align}
		E_1=-\sqrt{\Delta^2+t_1^2}+\Delta',\,\,E_2=-\Delta,\,\,E_3=\Delta,\,\,E_4=\sqrt{\Delta^2+t_1^2}+\Delta',
	\end{align}
	corresponding to the following eigenstates
	\begin{align}\label{Eq:BBG_wavefcn}
		\psi_1\approx\frac{1}{\sqrt{2}}\left[\begin{array}{r}
			0\\
			1\\
			-1\\
			0
		\end{array}\right],\,	\psi_2=\left[\begin{array}{r}
			1\\
			0\\
			0\\
			0
		\end{array}\right],\,
		\psi_3=\left[\begin{array}{r}
			0\\
			0\\
			0\\
			1
		\end{array}\right],\,
		\psi_4\approx\frac{1}{\sqrt{2}}\left[\begin{array}{r}
			0\\
			1\\
			1\\
			0
		\end{array}\right],
	\end{align}
	where we have used $t_1\gg\Delta$. $E_2$ and $E_3$ correspond to the low-energy valence band edge and the low-energy conduction band edge respectively, while $E_1$ and $E_4$ are the high energy band edges.
	
	We are interested in a slightly hole-doped BBG with a sufficiently large $D>0$.
	Ignoring the trigonal wrapping terms, the zero energy states are precisely at $K$ and $-K$ valleys, and $\hat{h}_{\tau}(\vex{k}\rightarrow 0)$ corresponds to a collection of decoupled atomic sites in the position space. Since $\hat{h}_{+}(0)=\hat{h}_{-}(0)$, we can treat the valley as an internal quantum number, and we construct an effective honeycomb lattice. 
	In such a model, we concentrate only on 1A and 1B sites with $\vex{k}\approx 0$. The minimal model for BBG contains only the $E_2$ and $E_4$ bands because contributions form other bands are much weaker. We explain the relevant band degrees of freedom in the following.
	
	First, $\psi_2$ has weights only on the 1A sites, so the $E_2$ band must be included for the interlayer tunneling.
	On the contrary, the $E_3$ band can be ignored because $\psi_3$ has weights only on the 2B sites. 
	Since Both $\psi_1$ and $\psi_4$ have finite weights on the 1B sites, we need to investigate the virtual processes involving both the $E_1$ and $E_4$ bands. 
	The primary interlayer process involving the $E_1$ band corresponds to tunneling an electron from the BBG $E_1$ band to the WSe$_2$ conduction bands; the primary interlayer process involving the $E_4$ band corresponds to tunneling an electron from the WSe$_2$ valence bands to the BBG $E_4$ band. For the induced ISOC, the spin-splitting in the WSe$_2$ conduction bands is much smaller than the valence bands, so we can ignore the contributions from the WSe$_2$ conduction bands. For the effective attraction in hole-doped BBG, the virtual process (i), involving the BBG $E_1$ band and the WSe$_2$ conduction bands, is a much weaker than the virtual process (ii), involving the BBG $E_4$ band and the WSe$_2$ valence bands. The results are due to the Coulomb interaction in the virtual states such that the energy is increased by $2U_1$ in (i) and the energy is decreased by $2U_1$ in (ii). Thus, the process (i) yields a much smaller contribution, and we focus only on the process (ii).
	Therefore, the interlayer tunneling processes with the $E_2$ and $E_4$ bands are the dominating contributions.
	
	The effective honeycomb lattice in the main text is related to the top graphene layer but with some differences. The B sites of the effective honeycomb model are \textit{not} the microscopic BBG 1B sites because we ignore the $E_1$ band completely. In the main text, we consider $E_A=0$ and $E_B=\sqrt{t_1^2+\Delta^2}+\Delta+\Delta'$ (the energy difference between $E_4$ and $E_2$ bands). In addition, the charge neutrality configuration in the effective model is described by completely filled A sites and empty B sites, which is quite different from the microscopic graphene where both sites have similar electron density. This effective honeycomb lattice description allows us to present the effective attraction in an intuitive and transparent way.
	
	\section{Monolayer Tungsten Diselenide}	
	
	The WSe$_2$ monolayer can be modeled by a massive Dirac model \cite{Xiao2012S} given by
	\begin{align}\label{Eq:h_k_WSe2}
		\hat{h}_{\tau}^{(\text{WSe}_2)}(\vex{k})=v\left(\tau\hat{\sigma}_xk_x+\hat{\sigma}_yk_y\right)+\frac{\Omega}{2}\hat{\sigma}_z+\lambda\tau\frac{\hat{\sigma}_z-\hat{1}}{2}\hat{s}_z=\left[\begin{array}{cccc}
			\Omega/2 & 0 & v(\tau k_x-ik_y) & 0\\
			0 & \Omega/2 & 0 & v(\tau k_x-ik_y)\\
			v(\tau k_x-ik_y) & 0 & -\Omega/2+\lambda & 0 \\
			0 & v(\tau k_x-ik_y) & 0 & -\Omega/2-\lambda
		\end{array}
		\right]
	\end{align}
	where $v=td$, $d=3.31$\AA, $t=1.19$eV, $\Omega=1.6$eV and $2\lambda=0.46$eV. The basis of $\hat{h}_{\tau}^{(\text{WSe}_2)}(\vex{k})$ is (C$\uparrow$, C$\downarrow$, V$\uparrow$, V$\downarrow$), where The label C corresponds to the $d_{z^2}$ orbital function and the label V corresponds to the $d_{x^2-y^2}+i\tau d_{xy}$ orbital function \cite{Xiao2012S}. 
	The eigenvalues of $\hat{h}_{\tau}^{(\text{WSe}_2)}(\vex{k})$ are given by 
	\begin{subequations}
		\begin{align}
			E^{(\text{WSe}_2)}_1(\vex{k})=&\frac{-\lambda-\sqrt{4v^2\vex{k}^2+(\Omega+\lambda)^2}}{2},\\
			E^{(\text{WSe}_2)}_2(\vex{k})=&\frac{\lambda-\sqrt{4v^2\vex{k}^2+(\Omega-\lambda)^2}}{2},\\
			E^{(\text{WSe}_2)}_3(\vex{k})=&\frac{-\lambda+\sqrt{4v^2\vex{k}^2+(\Omega+\lambda)^2}}{2},\\
			E^{(\text{WSe}_2)}_4(\vex{k})=&\frac{\lambda+\sqrt{4v^2\vex{k}^2+(\Omega-\lambda)^2}}{2},				
		\end{align}
	\end{subequations}
	corresponding to
	\begin{subequations}
		\begin{align}\label{Eq:WSe2_wavefcn}
			\psi_1^{(\text{WSe}_2)}(\vex{k})=&\frac{1}{\sqrt{1+\left[\frac{2v|\vex{k}|}{\Omega+\lambda+\sqrt{4v^2\vex{k}^2+(\Omega+\lambda)^2}}\right]^2}}\left[\begin{array}{c}
				0\\
				\frac{2v(-k_x+ik_y)}{\Omega+\lambda+\sqrt{4v^2\vex{k}^2+(\Omega+\lambda)^2}}\\
				0\\
				1
			\end{array}\right]\approx\frac{1}{\sqrt{1+\frac{v^2\vex{k}^2}{(\Omega+\lambda)^2}}}\left[\begin{array}{c}
				0\\
				\frac{v(-k_x+ik_y)}{\Omega+\lambda}\\
				0\\
				1
			\end{array}\right],\\	
			\psi_2^{(\text{WSe}_2)}(\vex{k})=&\frac{1}{\sqrt{1+\left[\frac{2v|\vex{k}|}{\Omega-\lambda+\sqrt{4v^2\vex{k}^2+(\Omega-\lambda)^2}}\right]^2}}\left[\begin{array}{c}
				\frac{2v(-k_x+ik_y)}{\Omega-\lambda+\sqrt{4v^2\vex{k}^2+(\Omega-\lambda)^2}}\\
				0\\
				1\\
				0
			\end{array}\right]\approx\frac{1}{\sqrt{1+\frac{v^2\vex{k}^2}{(\Omega-\lambda)^2}}}\left[\begin{array}{c}
				\frac{v(-k_x+ik_y)}{\Omega-\lambda}\\
				0\\
				1\\
				0
			\end{array}\right],\\
			\psi_3^{(\text{WSe}_2)}(\vex{k})=&\frac{1}{\sqrt{1+\left[\frac{2v|\vex{k}|}{\Omega+\lambda+\sqrt{4v^2\vex{k}^2+(\Omega+\lambda)^2}}\right]^2}}\left[\begin{array}{c}
				0\\
				1\\
				0\\
				\frac{2v(k_x+ik_y)}{\Omega+\lambda+\sqrt{4v^2\vex{k}^2+(\Omega+\lambda)^2}}
			\end{array}\right]\approx\frac{1}{\sqrt{1+\frac{v^2\vex{k}^2}{(\Omega+\lambda)^2}}}\left[\begin{array}{c}
				0\\
				1\\
				0\\
				\frac{v(k_x+ik_y)}{\Omega+\lambda}
			\end{array}\right],\\
			\psi_4^{(\text{WSe}_2)}(\vex{k})=&\frac{1}{\sqrt{1+\left[\frac{2v|\vex{k}|}{\Omega-\lambda+\sqrt{4v^2\vex{k}^2+(\Omega-\lambda)^2}}\right]^2}}\left[\begin{array}{c}
				1\\
				0\\
				\frac{2v(k_x+ik_y)}{\Omega-\lambda+\sqrt{4v^2\vex{k}^2+(\Omega-\lambda)^2}}\\
				0
			\end{array}\right]\approx\frac{1}{\sqrt{1+\frac{v^2\vex{k}^2}{(\Omega-\lambda)^2}}}\left[\begin{array}{c}
				1\\
				0\\
				\frac{v(k_x+ik_y)}{\Omega-\lambda}\\
				0
			\end{array}\right].
		\end{align}			
	\end{subequations}
	In the above expressions, we have assumed that $v|\vex{k}|\ll\Omega-\lambda<\Omega+\lambda$. At $\vex{k}=0$, $E_1^{\text{WSe}_2}=-\Omega/2-\lambda$, $E_2^{\text{WSe}_2}=-\Omega/2+\lambda$, and $E_3^{\text{WSe}_3}=E_4^{\text{WSe}_3}=\Omega/2$. Assuming that WSe$_2$ layer has a potential energy lower than BBG top layer by $\Delta$ (tuned by the displacement field), we can identify that $W=\Delta+\Omega/2-\lambda$ and $\delta=2\lambda$.
	
	\section{Derivation of ISOC and virtual-tunneling-induced attraction}
	
	The effective model in the main text allows for an intuitive understanding on ISOC as well as virtual-tunneling-induced attraction by treating interlayer tunneling using second-order perturbation theory. The unperturbed Hamiltonian is described by $\hat{H}_0=\hat{H}_{\text{tG}}+\hat{H}_I+\hat{H}_d$, where $\hat{H}_{\text{tG}}$ is the effective honeycomb lattice, $\hat{H}_I$ encodes the short-range Coulomb interaction, and $\hat{H}_d$ represents the effective energy levels for WSe$_2$. The perturbed Hamiltonian is the interlayer tunneling given by $\hat{H}_V$. Note that the ground state of the unperturbed Hamiltonian is described by filled WSe$_2$ levels and dilute (singly occupied) holes on A sites of the effective honeycomb lattice model. The correction of energy in second-order perturbation theory is given by
	\begin{align}
		\delta E_g=\sum_e\langle g|\hat{H}_V|e\rangle\langle e|\frac{1}{E_g-\hat{H}_{0}}|e\rangle\langle e|\hat{H}_V\hat{P}|g\rangle,
	\end{align}
	where $E_g$ and $|g\rangle$ are the energy and the wavefunction of the ground state of $\hat{H}_0$, $|e\rangle$ denotes the excited state of $\hat{H}_0$, and the sum runs over the excited states of $\hat{H}_0$. The ground state is described by dilute holes on A sites, and no more than two holes are occupied at the same position. Since the Fermi energy is at the band edge of the valence band, the ground states consists a few holes. Specifically, we focus on one-hole states and two-hole states (with two holes connected by a nearest-neighbor common B site).
	
	\subsection{Ising spin-orbit coupling}
	
	To derive ISOC, we consider single hole ground states and compute the correction of energies based of the valley and spin quantum numbers. The energy correction to a hole with valley $\tau$ and spin $s$ is given by $E_{\tau s}$, where
	\begin{align}
		\delta E_{+\uparrow}=&-\frac{\left|V^A_{+\uparrow}\right|^2}{W}-\frac{\left|\bar{V}^A_{-\uparrow}\right|^2}{W+\delta},\\
		\delta E_{+\downarrow}=&-\frac{\left|V^A_{+\downarrow}\right|^2}{W+\delta}-\frac{\left|\bar{V}^A_{-\downarrow}\right|^2}{W},\\
		\delta E_{-\uparrow}=&-\frac{\left|V^A_{-\uparrow}\right|^2}{W+\delta}-\frac{\left|\bar{V}^A_{+\uparrow}\right|^2}{W},\\
		\delta E_{-\downarrow}=&-\frac{\left|V^A_{-\downarrow}\right|^2}{W}-\frac{\left|\bar{V}^A_{+\downarrow}\right|^2}{W+\delta}.
	\end{align}
	Since $\left|V^A_{\tau s}\right|=\left|V^A_{\bar{\tau}\bar{s}}\right|$ and $\left|\bar{V}^A_{\tau s}\right|=\left|\bar{V}^A_{\bar{\tau}\bar{s}}\right|$, we can show that $\delta E_{\tau s}=E_{\bar{\tau}\bar{s}}$, realizing ISOC. The ISOC parameter is defined by
	\begin{align}\label{Eq:SM:lambda_I}
		\lambda_I\equiv-\delta E_{+\uparrow}+\delta E_{+\downarrow}=\frac{\left|V^A_{+\uparrow}\right|^2}{W}+\frac{\left|\bar{V}^A_{-\uparrow}\right|^2}{W+\delta}-\frac{\left|V^A_{+\downarrow}\right|^2}{W+\delta}-\frac{\left|\bar{V}^A_{-\downarrow}\right|^2}{W}=\frac{\left|V^A_{+\uparrow}\right|^2-\left|\bar{V}^A_{+\uparrow}\right|^2}{W}
		-\frac{\left|V^A_{+\downarrow}\right|^2-\left|\bar{V}^A_{+\downarrow}\right|^2}{W+\delta}
	\end{align}		
	
	\subsection{Virtual-tunneling-induced attraction}

\begin{figure}[t!]
		\includegraphics[width=0.4\textwidth]{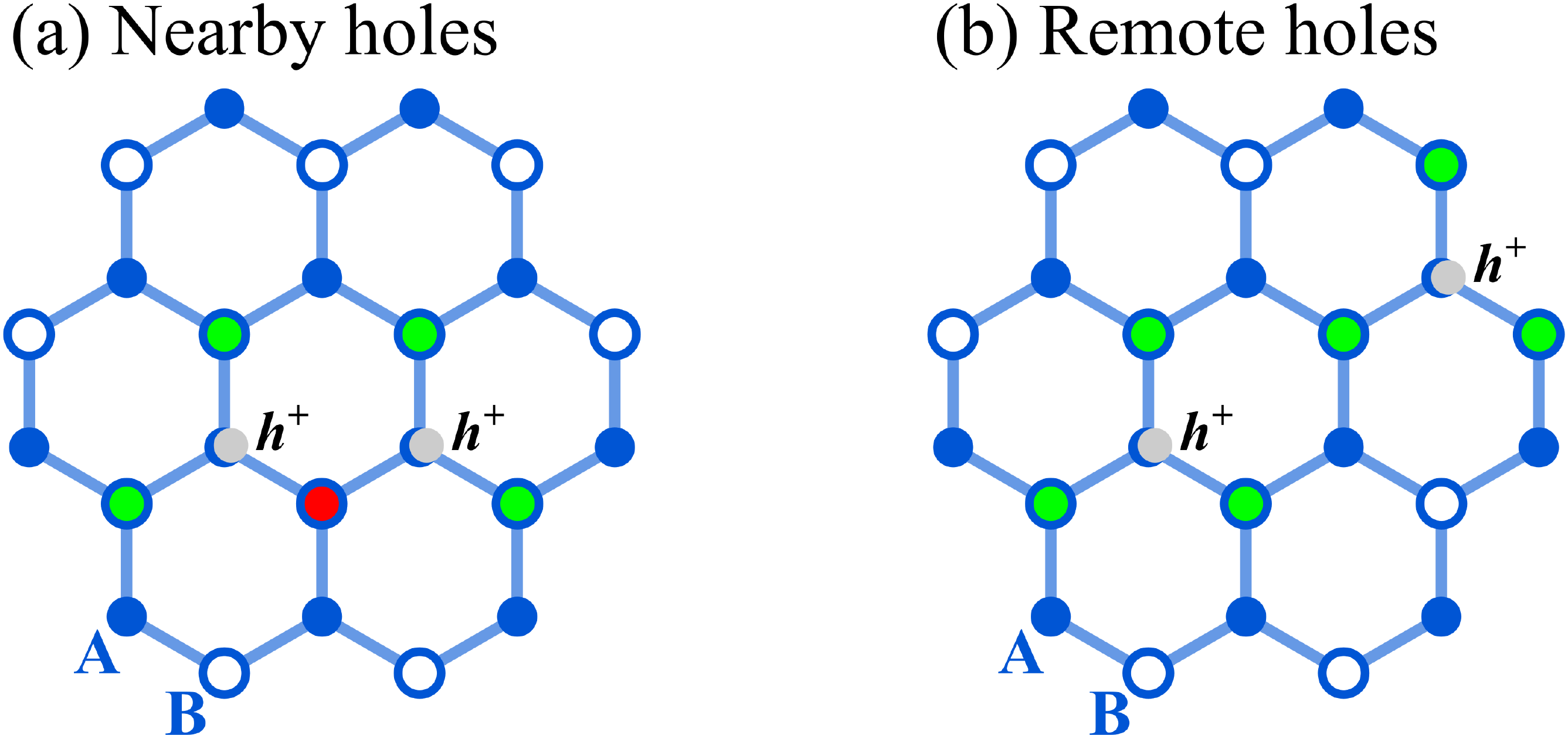}
		\caption{Configurations of two-hole states. (a) Two nearby holes that are connected by a common B site. (b) Two remote holes. The virtual tunneling induced attraction involves B sites, and we need to compare the energy difference between (a) and (b). The white circles correspond to the on-site energy correction $E_v^{(0)}$ (without any nearest-neighbor hole), the green circles correspond to the on-site energy correction $E_v^{(1)}$ (with one nearest-neighbor hole), the red circles correspond to the on-site energy correction $E_v^{(2)}$ (with two nearest-neighbor holes). 
		}		
		\label{Fig:E_difference}
	\end{figure}

	Now, we discuss the effective attraction induced by virtual tunnelings. We consider a state with two holes that are located at two nearest-neighbor A sites as illustrated in Fig.~2(c) in the main text. Two A sites are connected by a common B site (marked by $\vex{x}_B$), which represents a high energy conduction band of BBG. We compare the energy difference between the state with two nearby holes and the state with two far separated holes. Note that we need to take into account the entire energy renormalization due to the virtual interlayer tunneling. In Fig.~\ref{Fig:E_difference}, we plot two configurations of states, and we use different colors to specify the different virtual tunneling contributions. Specifically, the white circles correspond to the on-site energy correction $E_v^{(0)}$ (without any nearest-neighbor hole), the green circles correspond to the on-site energy correction $E_v^{(1)}$ (with one nearest-neighbor hole), the red circles correspond to the on-site energy correction $E_v^{(2)}$ (with two nearest-neighbor holes). These onsite energy correction at the second order are given by
	\begin{align}
		E_v^{(0)}=&-\frac{\left|V^{B}_{+\uparrow}\right|^2+\left|\bar{V}^{B}_{+\uparrow}\right|^2+\left|V^{B}_{-\downarrow}\right|^2+\left|\bar{V}^{B}_{-\downarrow}\right|^2}{W+E_B}-\frac{\left|V^{B}_{-\uparrow}\right|^2+\left|\bar{V}^{B}_{-\uparrow}\right|^2+\left|V^{B}_{+\downarrow}\right|^2+\left|\bar{V}^{B}_{+\downarrow}\right|^2}{W+\delta+E_B},\\
		E_v^{(1)}=&-\frac{\left|V^{B}_{+\uparrow}\right|^2+\left|\bar{V}^{B}_{+\uparrow}\right|^2+\left|V^{B}_{-\downarrow}\right|^2+\left|\bar{V}^{B}_{-\downarrow}\right|^2}{W+E_B-U_1}-\frac{\left|V^{B}_{-\uparrow}\right|^2+\left|\bar{V}^{B}_{-\uparrow}\right|^2+\left|V^{B}_{+\downarrow}\right|^2+\left|\bar{V}^{B}_{+\downarrow}\right|^2}{W+\delta+E_B-U_1},\\
		E_v^{(2)}=&-\frac{\left|V^{B}_{+\uparrow}\right|^2+\left|\bar{V}^{B}_{+\uparrow}\right|^2+\left|V^{B}_{-\downarrow}\right|^2+\left|\bar{V}^{B}_{-\downarrow}\right|^2}{W+E_B-2U_1}-\frac{\left|V^{B}_{-\uparrow}\right|^2+\left|\bar{V}^{B}_{-\uparrow}\right|^2+\left|V^{B}_{+\downarrow}\right|^2+\left|\bar{V}^{B}_{+\downarrow}\right|^2}{W+\delta+E_B-2U_1}.
	\end{align}
	The energy difference between the nearby-holes state and the remote-holes state is given by
	\begin{align}
		\Delta E=&E_v^{(0)}+E_v^{(2)}-2E_v^{(1)}\\
		\nonumber=&-\left(2\left|V^{B}_{+\uparrow}\right|^2+2\left|\bar{V}^{B}_{+\uparrow}\right|^2\right)\left[\frac{1}{W+E_B}+\frac{1}{W+E_B-2U_1}-\frac{2}{W+E_B-U_1}\right]\\
		&-\left(2\left|V^{B}_{+\downarrow}\right|^2+2\left|\bar{V}^{B}_{+\downarrow}\right|^2\right)\left[\frac{1}{W+\delta+E_B}+\frac{1}{W+\delta+E_B-2U_1}-\frac{2}{W+\delta+E_B-U_1}\right],
	\end{align}
	where we have used $\left|V^B_{\tau s}\right|=\left|V^B_{\bar{\tau}\bar{s}}\right|$ and $\left|\bar{V}^B_{\tau s}\right|=\left|\bar{V}^B_{\bar{\tau}\bar{s}}\right|$.
	Thus, the effective attraction in the main text is expressed by
	\begin{align}\label{Eq:SM:U_eff}
		\nonumber\mathcal{U}_{\text{eff}}=&\left(2\left|V^{B}_{+\uparrow}\right|^2+2\left|\bar{V}^{B}_{+\uparrow}\right|^2\right)\left[\frac{1}{W+E_B}+\frac{1}{W+E_B-2U_1}-\frac{2}{W+E_B-U_1}\right]\\
		&+\left(2\left|V^{B}_{+\downarrow}\right|^2+2\left|\bar{V}^{B}_{+\downarrow}\right|^2\right)\left[\frac{1}{W+\delta+E_B}+\frac{1}{W+\delta+E_B-2U_1}-\frac{2}{W+\delta+E_B-U_1}\right].
	\end{align}
	We note that $\mathcal{U}_{\text{eff}}\propto U_1^2$ for an infinitesimal $U_1$, indicating that the effective attraction is absent in the noninteracting limit.

	\subsection{Stability of second-order perturbation theory}
	
	Since our analysis is based on the second-order perturbation, it is important to justify that the higher-order contributions are small. The expansion parameters in our theory are $|V_A|/W$ and $|V_B|/(W+E_B - 2 U_1)$. Using $|V_A|\approx 100$meV (based on Ref.~\cite{David2019S}), we obtain $|V_A|/W\approx 0.16$ and $|V_B|/(W+E_B - 2 U_1)\approx 0.37$ for $U_1=0.43$eV ($V_B\approx V_A/\sqrt{2}$). The higher-order contributions start at the fourth order in tunneling, and there are two distinct contributions -- (a) fourth order in $V_B$ and (b) second order in $V_A$ and second order in $V_B$. The former case results in a quantitative change in $U_{\text{eff}}$, and the latter case generates an interaction that breaks spin symmetry. Using the naive counting estimate, the case (a) gives a correction of order $U_{\text{eff}}|V_B|^2/(W+E_B - 2 U_1)^2\approx 0.14U_{\text{eff}}$, and case (b) generates a spin-dependent interaction of order $U_{\text{eff}}|V_A|^2/W^2\approx 0.03U_{\text{eff}}$. Both contributions are small compared to $U_{\text{eff}}$, suggesting that the results based on second-order perturbation theory are reasonable.

	\section{Microscopic interlayer tunneling}	
	
	In the BBG-WSe$_2$ experiment, there is a relative angle $\theta$ between WSe$_2$ and BBG. In addition, the lattice constant of WSe$_2$ is $d=0.331$nm which is larger than the lattice constant of graphene, $a=0.246$nm. These two factors are crucial for the microscopic interlayer tunneling, which are discussed in the section.\\

\begin{figure}[t!]
		\includegraphics[width=0.45\textwidth]{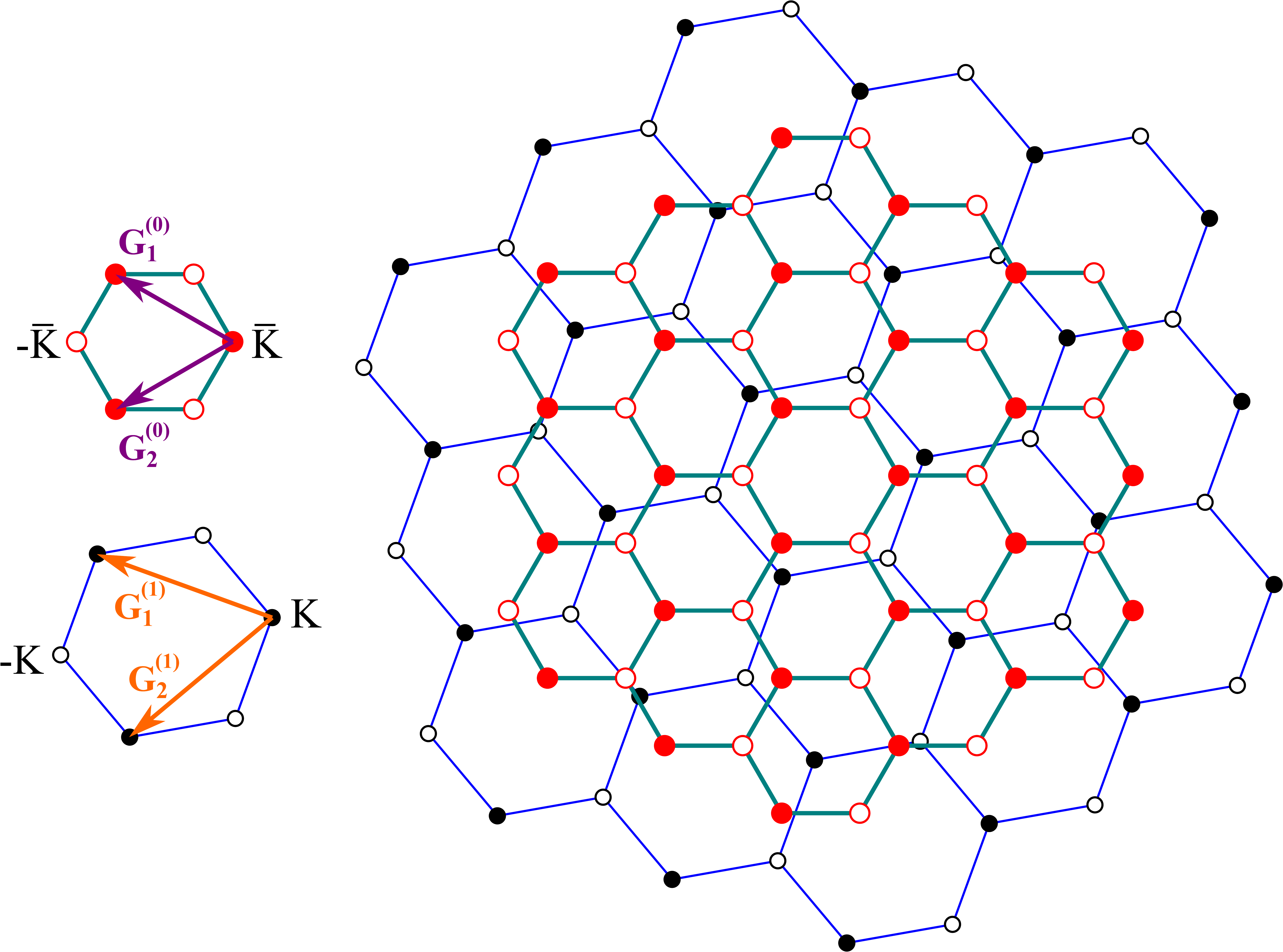}
		\caption{Brillouin zones of WSe$_2$ and BBG. The green (blue) lines indicate the reciprocal lattice of WSe$_2$ (BBG). The blue lattice is rotated by an angle $\theta=10^{\circ}$ in this illustration. $\vex{G}^{(0)}_1=\left(-\frac{2\pi}{d},\frac{2\pi}{\sqrt{3}d}\right)$, $\vex{G}^{(0)}_2=\left(-\frac{2\pi}{d},-\frac{2\pi}{\sqrt{3}d}\right)$, $\vex{G}^{(1)}_1=\frac{2\pi}{a}\left(-\cos\theta-\frac{\sin\theta}{\sqrt{3}},
			\frac{\cos\theta}{\sqrt{3}}-\sin\theta\right)$, $\vex{G}^{(1)}_2=\frac{2\pi}{a}\left(-\cos\theta+\frac{\sin\theta}{\sqrt{3}},
			-\frac{\cos\theta}{\sqrt{3}}-\sin\theta\right)$.
		}		
		\label{Fig:BZ}
	\end{figure}

	First of all, the interlayer tunneling preserves the crystal momentum because we assume that the tunneling between two positions depends only on their relative displacement. In our case, the Brillouin zones of WSe$_2$ and BBG are in different sizes, and WSe$_2$ is rotated by an angle $\theta$. In Fig.~\ref{Fig:BZ}, the Brillouin zones for both the WSe$_2$ and BBG are illustrated. 
	Within the two-center approximation scheme, the spin-preserving tunneling between two layers (from the $\beta$th band with sublattice $\sigma'$ to the $\alpha$ band with sublattice $\sigma$) can be described by \cite{Bistritzer2010S}
	\begin{align}\label{Eq:T_kp} T_{\vex{k},\vex{p}}^{\alpha,\sigma;\beta,\sigma'}\equiv T(\alpha,\sigma,\vex{k};\beta,\sigma',\vex{p})=&\frac{1}{\sqrt{\mathcal{A}_0\mathcal{A}_1}}\left(a^{(1)}_{\vex{k},\alpha,\sigma}\right)^*\left(a^{(0)}_{\vex{p},\beta,\sigma'}\right)\!\sum_{\vex{Q}^{(1)},\vex{Q}^{(0)}}\tilde{t}_{\vex{k}+\vex{Q}^{(1)}}e^{i\vex{Q}^{(1)}\cdot\vex{R}^{(1)}_{\sigma}}e^{-i\vex{Q}^{(0)}\cdot\vex{R}^{(0)}_{\sigma'}}\delta_{\vex{k}+\vex{Q}^{(1)},\vex{p}+\vex{Q}^{(0)}},
	\end{align}
	where $\mathcal{A}_0$ ($\mathcal{A}_1$) is unit cell area of the WSe$_2$ (graphene), $\alpha$ and $\beta$ are the band indexes, $\vex{k}$ and $\vex{p}'$ are the  momentum relative to $\Gamma$ point, $\sigma$ and $\sigma'$ are the sublattice indexes, $a^{(l)}_{\vex{k},\alpha,\sigma}$ is the sublattice projection of a Bloch state $|\vex{k},\alpha\rangle$ at layer $l$, $\vex{R}^{(l)}_{\sigma}$ is the sublattice vector of $l$th layer, $\vex{Q}^{(0)}$ and $\vex{Q}^{(1)}$ represent the reciprocal lattice vectors in WSe$_2$ and BBG, respectively, and $\tilde{t}_{\vex{k}}$ is the 2D Fourier transform of the interlayer tunneling amplitude with a finite range. $T(\alpha,\sigma,\vex{k};\beta,\sigma',\vex{p})$ indicates the components of both bands and sublattices. In the above expression, we use index $0$ as the WSe$_2$ layer and index $1$ as the top graphene layer of BBG. We have ignored relative layer shift $\vex{d}$ for simplicity. $\vex{Q}^{(0)}=n_0\vex{G}^{(0)}_1+m_0\vex{G}^{(0)}_2$ and $\vex{Q}^{(1)}=n_1\vex{G}^{(1)}_1+m_1\vex{G}^{(1)}_2$ for integer values of $n_0$, $m_0$, $n_1$, $m_1$, and $\vex{G}^{(0)}_1$, $\vex{G}^{(0)}_2$, $\vex{G}^{(1)}_1$, $\vex{G}^{(1)}_2$ are the primitive lattice vectors for the reciprocal lattices as illustrated in Fig.~\ref{Fig:BZ}. $\vex{R}^{(0)}_{A}=(0,0)$, $\vex{R}^{(0)}_{B}=(0,d/\sqrt{3})$, $\vex{R}^{(1)}_{A}=(0,0)$, and $\vex{R}^{(1)}_{B}=(0,a/\sqrt{3})$.\\
	
	Equation~(\ref{Eq:T_kp}) describes the microscopic interlayer tunneling between two layers. The contributions come from infinitely many crystal momenta, but the contributions from the large crystal momentum $\vex{k}$ is suppressed by $\tilde{t}_{\vex{k}}$, which can be modeled as a stretched exponentially decaying function in $\vex{k}$ \cite{Bistritzer2010S} (will be discussed later). In addition, the contribution is small if the momentum $\vex{k}$ is too far away from the valleys due to the structure of $a^{(l)}_{\vex{k},\alpha,\sigma}$.\\
	
	Our main purpose is to connect the microscopic calculations to the minimal model approach. We focus only on the tunneling between the $d_{x^2-y^2}\pm i d_{xy}$ orbitals of WSe$_2$ and the top graphene layer of BBG. Since our theory is based on electrons tunneling from WSe$_2$ valence bands, only the 1st and 2nd bands in $\hat{h}_{\tau}^{(\text{WSe}_2)}(\vex{k})$ [given by Eq.~(\ref{Eq:h_k_WSe2})] are considered. With respect to BBG, only the 2nd and 4th bands in $\hat{h}_{\tau}(\vex{k})$ [given by Eq.~(\ref{Eq:h_k_BBG})] are considered because these two bands are relevant to the intervalley tunneling in our model. To simplify our calculations, we keep only $K$ and $-K$ momenta in BBG bands, and we consider a momentum cutoff $\bar\Lambda$ (relative to $\bar{K}$ or $-\bar{K}$) for WSe$_2$ bands. We define $\vex{K}_0=\frac{4\pi}{3 d}(1,0)$ and $\vex{K}_1=\frac{4\pi}{3 a}(\cos\theta,\sin\theta)$, where $a=0.246$nm (lattice constant of graphene) and $d=0.331$nm (lattice constant of WSe$_2$). 
	
	\subsection{ISOC and virtual-tunneling-induced attraction}
	
	In the main text, we discuss ideas of deriving ISOC and the virtual-hopping-induced attraction. The same procedures can be done with the microscopic interlayer tunneling. The main differences is that multiple momenta states in WSe$_2$ can contribute to the virtual processes instead of one. This is a consequence of the moir\'e structure of the BBG-WSe$_2$ such that crystal momentum conservation can allow for tunnelings from extended zones.\\
	
	We can formally compute ISOC based on second order perturbation theory of the microscopic interlayer tunneling and derive that
	\begin{align}\label{Eq:lambda_I_micro}
		\lambda_I=\sum_{\vex{q},|\vex{q}|<\bar\Lambda}\left[\begin{array}{c}
			\frac{\left|T(2,A,\vex{K}_1;2,B,\vex{K}_0+\vex{q})\right|^2}{-E_2^{(\text{WSe}_2)}(\vex{q})+2\Delta}-\frac{\left|T(2,A,\vex{K}_1;1,B,\vex{K}_0+\vex{q})\right|^2}{-E_1^{(\text{WSe}_2)}(\vex{q})+2\Delta}-\frac{\left|T(2,A,-\vex{K}_1;2,B,\vex{K}_0+\vex{q})\right|^2}{-E_2^{(\text{WSe}_2)}(\vex{q})+2\Delta}+\frac{\left|T(2,A,-\vex{K}_1;1,B,\vex{K}_0+\vex{q})\right|^2}{-E_1^{(\text{WSe}_2)}(\vex{q})+2\Delta}
		\end{array}\right],
	\end{align}
	where we have incorporated the potential energy difference ($2\Delta$) between WSe$_2$ and the top graphene layer (assuming the same distance as the top and bottom graphene layers) and $\bar{\Lambda}$ is the cutoff for the momentum sum. $\bar\Lambda\le\frac{2\pi}{3d}$ is required for consistently defining valleys. In this work, we use $\bar\Lambda=\frac{2\pi}{3d}=0.633$\AA$^{-1}$. With the above expression and Eq.~(\ref{Eq:SM:lambda_I}), we identify that
	\begin{subequations}\label{Eq:V_hopping_ISOC}
		\begin{align}
			\frac{\left|V^{A}_{+\uparrow}\right|^2}{W}=&\sum_{\vex{q},|\vex{q}|<\bar\Lambda}\frac{\left|T(2,A,\vex{K}_1;2,B,\vex{K}_0+\vex{q})\right|^2}{-E_2^{(\text{WSe}_2)}(\vex{q})+2\Delta},\\
			\frac{\left|V^{A}_{+\downarrow}\right|^2}{W+\delta}=&\sum_{\vex{q},|\vex{q}|<\bar\Lambda}\frac{\left|T(2,A,\vex{K}_1;1,B,\vex{K}_0+\vex{q})\right|^2}{-E_1^{(\text{WSe}_2)}(\vex{q})+2\Delta},\\
			\frac{\left|\bar{V}^{A}_{+\uparrow}\right|^2}{W}=&\sum_{\vex{q},|\vex{q}|<\bar\Lambda}\frac{\left|T(2,A,-\vex{K}_1;2,B,\vex{K}_0+\vex{q})\right|^2}{-E_2^{(\text{WSe}_2)}(\vex{q})+2\Delta},\\
			\frac{\left|\bar{V}^{A}_{+\downarrow}\right|^2}{W+\delta}=&\sum_{\vex{q},|\vex{q}|<\bar\Lambda}\frac{\left|T(2,A,-\vex{K}_1;1,B,\vex{K}_0+\vex{q})\right|^2}{-E_1^{(\text{WSe}_2)}(\vex{q})+2\Delta}.
		\end{align}
	\end{subequations}
	
	Similarly, we can compute the virtual-tunneling-induced attraction and derive that
	\begin{align}\label{Eq:U_eff_micro}
		\mathcal{U}_{\text{eff}}=2\sum_{\vex{q},|\vex{q}|<\bar\Lambda}\left[\begin{array}{c}
			\frac{\left|T(4,B,\vex{K}_1;2,B,\vex{K}_0+\vex{q})\right|^2}{-E_2^{(\text{WSe}_2)}(\vex{q})+2\Delta+E_B}+\frac{\left|T(4,B,-\vex{K}_1;2,B,\vex{K}_0+\vex{q})\right|^2}{-E_2^{(\text{WSe}_2)}(\vex{q})+2\Delta+E_B}+\frac{\left|T(4,B,\vex{K}_1;1,B,\vex{K}_0+\vex{q})\right|^2}{-E_1^{(\text{WSe}_2)}(\vex{q})+2\Delta+E_B}+\frac{\left|T(4,B,-\vex{K}_1;1,B,\vex{K}_0+\vex{q})\right|^2}{-E_1^{(\text{WSe}_2)}(\vex{q})+2\Delta+E_B}\\[4mm]
			+\frac{\left|T(4,B,\vex{K}_1;2,B,\vex{K}_0+\vex{q})\right|^2}{-E_2^{(\text{WSe}_2)}(\vex{q})+2\Delta+E_B-2U_1}+\frac{\left|T(4,B,-\vex{K}_1;2,B,\vex{K}_0+\vex{q})\right|^2}{-E_2^{(\text{WSe}_2)}(\vex{q})+2\Delta+E_B-2U_1}+\frac{\left|T(4,B,\vex{K}_1;1,B,\vex{K}_0+\vex{q})\right|^2}{-E_1^{(\text{WSe}_2)}(\vex{q})+2\Delta+E_B-2U_1}+\frac{\left|T(4,B,-\vex{K}_1;1,B,\vex{K}_0+\vex{q})\right|^2}{-E_1^{(\text{WSe}_2)}(\vex{q})+2\Delta+E_B-2U_1}\\[4mm]
			-2\frac{\left|T(4,B,\vex{K}_1;2,B,\vex{K}_0+\vex{q})\right|^2}{-E_2^{(\text{WSe}_2)}(\vex{q})+2\Delta+E_B-U_1}-2\frac{\left|T(4,B,-\vex{K}_1;2,B,\vex{K}_0+\vex{q})\right|^2}{-E_2^{(\text{WSe}_2)}(\vex{q})+2\Delta+E_B-U_1}-2\frac{\left|T(4,B,\vex{K}_1;1,B,\vex{K}_0+\vex{q})\right|^2}{-E_1^{(\text{WSe}_2)}(\vex{q})+2\Delta+E_B-U_1}-2\frac{\left|T(4,B,-\vex{K}_1;1,B,\vex{K}_0+\vex{q})\right|^2}{-E_1^{(\text{WSe}_2)}(\vex{q})+2\Delta+E_B-U_1}
		\end{array}
		\right].
	\end{align}
	With the above expression and Eq~(\ref{Eq:SM:U_eff}), we identify that
	\begin{subequations}\label{Eq:V_hopping_att}
		\begin{align}
			\frac{\left|V^{B}_{+\uparrow}\right|^2}{W+E_B-xU_1}=&\sum_{\vex{q},|\vex{q}|<\bar\Lambda}\frac{\left|T(4,B,\vex{K}_1;2,B,\vex{K}_0+\vex{q})\right|^2}{-E_2^{(\text{WSe}_2)}(\vex{q})+2\Delta+E_B-xU_1},\\
			\frac{\left|V^{B}_{+\downarrow}\right|^2}{W+\delta+E_B-xU_1}=&\sum_{\vex{q},|\vex{q}|<\bar\Lambda}\frac{\left|T(4,B,\vex{K}_1;1,B,\vex{K}_0+\vex{q})\right|^2}{-E_1^{(\text{WSe}_2)}(\vex{q})+2\Delta+E_B-xU_1},\\
			\frac{\left|\bar{V}^{B}_{+\uparrow}\right|^2}{W+E_B-xU_1}=&\sum_{\vex{q},|\vex{q}|<\bar\Lambda}\frac{\left|T(4,B,-\vex{K}_1;2,B,\vex{K}_0+\vex{q})\right|^2}{-E_2^{(\text{WSe}_2)}(\vex{q})+2\Delta+E_B-xU_1},\\
			\frac{\left|\bar{V}^{B}_{+\downarrow}\right|^2}{W+\delta+E_B-xU_1}=&\sum_{\vex{q},|\vex{q}|<\bar\Lambda}\frac{\left|T(4,B,-\vex{K}_1;1,B,\vex{K}_0+\vex{q})\right|^2}{-E_1^{(\text{WSe}_2)}(\vex{q})+2\Delta+E_B-xU_1},
		\end{align}
	\end{subequations}
	where $x=0,1,2$.
	
	\subsection{Evaluating interlayer tunneling}

	To compute $T(\alpha,\sigma,\vex{k};\beta,\sigma',\vex{p})$, $a^{(l)}_{\vex{k},\alpha,\sigma}$ and $\tilde{t}_{\vex{k}}$ are needed. 
	$a^{(l)}_{\vex{k},\alpha,\sigma}$ corresponds to the wavefunctions of $\hat{h}_{\tau}^{(\text{WSe}_2)}(\vex{k})$ and $\hat{h}_{\tau}(\vex{k})$, given by Eqs.~(\ref{Eq:h_k_WSe2}) and (\ref{Eq:BBG_wavefcn}). We find
	\begin{align}
		a^{(0)}_{\vex{K}_0+\vex{q},1,B}=\frac{1}{\sqrt{1+\left[\frac{2v|\vex{q}|}{\Omega+\lambda+\sqrt{4v^2\vex{q}^2+(\Omega+\lambda)^2}}\right]^2}},\,\,
		a^{(0)}_{\vex{K}_0+\vex{q},2,B}=\frac{1}{\sqrt{1+\left[\frac{2v|\vex{q}|}{\Omega-\lambda+\sqrt{4v^2\vex{q}^2+(\Omega-\lambda)^2}}\right]^2}},\,\,
		a^{(1)}_{\pm \vex{K}_1,2,A}=1,\,\,
		a^{(1)}_{\pm \vex{K}_1,4,B}=\frac{1}{\sqrt{2}}.
	\end{align}
	Based on the wavefunction amplitudes above, we conclude that $\left|T(2,A,\vex{K}_1;2,B,\vex{K}_0+\vex{q})\right|<\left|T(2,A,\vex{K}_1;1,B,\vex{K}_0+\vex{q})\right|$ and $\left|T(4,B,\vex{K}_1;2,B,\vex{K}_0+\vex{q})\right|<\left|T(4,B,\vex{K}_1;1,B,\vex{K}_0+\vex{q})\right|$, but the difference might be small. In addition, $\left|T(4,B,\vex{K}_1;1,B,\vex{K}_0+\vex{q})\right|=\frac{1}{\sqrt{2}}\left|T(2,A,\vex{K}_1;1,B,\vex{K}_0+\vex{q})\right|$ and $\left|T(4,B,\vex{K}_1;2,B,\vex{K}_0+\vex{q})\right|=\frac{1}{\sqrt{2}}\left|T(2,A,\vex{K}_1;2,B,\vex{K}_0+\vex{q})\right|$.\\
	
	Regarding $\tilde{t}_{\vex{k}}$, we follow Ref.~\cite{Bistritzer2010S} and adopt a stretched exponential ansatz as follows:
	\begin{align}
		|\tilde{t}_{\vex{k}}|=t_0 e^{-\chi(|\vex{k}|z_{\perp})^{\gamma}},
	\end{align}
	where $t_0$ is an overall constant, $\chi$ is an order 1 numerical constant, $\gamma$ is the exponent of stretched exponential, and $z_{\perp}$ is the distance between WSe$_2$ monolayer and the top graphene layer of BBG. We use $z_{\perp}=3.34${\AA }, the same as the distance between two graphene layers. $\chi$ and $\gamma$ control the decay of $\tilde{t}_{\vex{k}}$ for a sufficiently large $\vex{k}$, and we will discuss the results with a few representative values of $\chi$ and $\gamma$ in addition to the twist angle $\theta$. 
	Since $\tilde{t}_{\vex{k}}$ becomes small for a sufficiently large $\vex{k}$, we can ignore contributions such that $\vex{k}+\vex{Q}^{(1)}>\Lambda$ in Eq.~(\ref{Eq:T_kp}), where we use $\Lambda=\frac{4\pi}{a}=5.11$\AA$^{-1}$.\\

	\begin{figure}[t!]
		\includegraphics[width=0.8\textwidth]{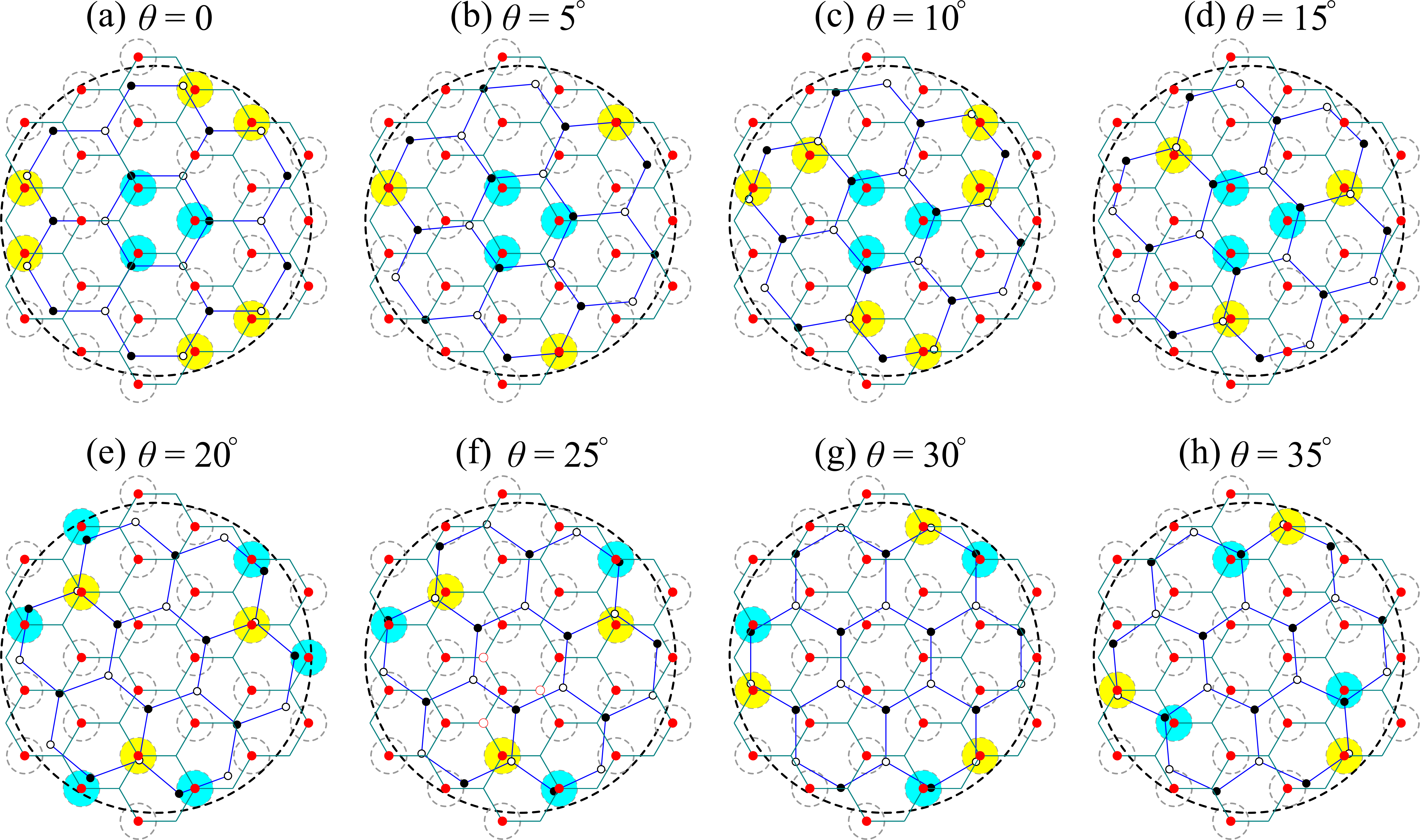}
		\caption{Kinematics of tunneling matrix elements with different angles. The green and the blue lines draw the Brillouin zone boundaries of WSe$_2$ and BBG, respectively. The gray circles with dashed lines indicate the cutoff $\bar\Lambda$ of the momentum relative to the $\bar{K}$ valleys (marked by red solid dots) of WSe$_2$. The big black circle with dashed lines indicates the cutoff $\Lambda$ of the total momentum (relative to $\Gamma$ point in the first Brillouin zone). The blue regions and yellow regions indicate the finite contributions of intravalley and intervalley tunnelings, respectively. The black solid (opened) dots denote the $K$ ($-K$) valleys of BBG.}
		\label{Fig:BZ_tun}
	\end{figure}	
	
	In Fig.~\ref{Fig:BZ_tun}, the kinematics of momenta involved in tunneling is plotted for a few values of $\theta$. The big black dashed circle draws the region of $\vex{k}+\vex{Q}^{(1)}<\Lambda$ considered in the tunneling. We also use the small gray dashed circles to mark the momentum cutoff $\bar\Lambda$ for momentum relative to $\bar{K}$ valley. The blue circles indicate finite contributions of intravalley processes; the yellow circles indicate finite contributions of intervalley processes. Now, we are in the position to compute interlayer tunneling between WSe$_2$ and BBG.

	\subsection{ $T(\alpha,\sigma,k;\beta,\sigma',p)$ at different angles}
	
	The interlayer tunneling matrix element crucially depends on the twist angle $\theta$. In Fig.~\ref{Fig:BZ_tun}, we plot crystal momenta involved in the interlayer tunneling with different $\theta$. We will list all the needed tunneling matrix elements for $0\le\theta\le 30^{\circ}$. In our model, angle $60^{\circ}-\theta$ is related to $\theta$, but the intravalley and intervalley contributions are interchanged. See the kinematics in Figs.~\ref{Fig:BZ_tun}(f) and \ref{Fig:BZ_tun}(h) for an example.
	
	\subsubsection{$\theta=0$}
	
	For $\theta=0$, there are three momenta contributing to the intravalley tunnelings and six momenta contributing to the intervalley tunnelings. Since the three intravalley tunnelings are related by the $\mathcal{C}_{3z}$ rotation, we only need to evaluate one of them. The six intervalley tunneling momenta are also related by symmetry operation, so only one of them needs to be evaluated. The spin-dependent interlayer tunneling terms are as follows:
	\begin{subequations}
		\begin{align}
			\left|T(2,A,\vex{K}_1;2,B,\vex{K}_0+\vex{q})\right|^2\bigg|_{\vex{q}=\vex{K}_1-\vex{K}_0}\approx& \frac{t_0^2}{\mathcal{A}_0\mathcal{A}_1}\frac{e^{-2\chi(|\vex{K}_1|z_{\perp})^{\gamma}}}{1+\left[\frac{2v|\vex{K}_1-\vex{K}_0|}{\Omega-\lambda+\sqrt{4v^2|\vex{K}_1-\vex{K}_0|^2+(\Omega-\lambda)^2}}\right]^2},\\
			\left|T(2,A,\vex{K}_1;1,B,\vex{K}_0+\vex{q})\right|^2\bigg|_{\vex{q}=\vex{K}_1-\vex{K}_0}\approx&\frac{t_0^2}{\mathcal{A}_0\mathcal{A}_1}\frac{e^{-2\chi(|\vex{K}_1|z_{\perp})^{\gamma}}}{1+\left[\frac{2v|\vex{K}_1-\vex{K}_0|}{\Omega+\lambda+\sqrt{4v^2|\vex{K}_1-\vex{K}_0|^2+(\Omega+\lambda)^2}}\right]^2},\\
			\left|T(2,A,-\vex{K}_1;2,B,\vex{K}_0+\vex{q})\right|^2\bigg|_{\vex{q}=-\vex{K}_1-2\vex{G}^{(1)}_2-[\vex{K}_0+\vex{G}^{(0)}_1-2\vex{G}^{(0)}_2]}\approx&
			\frac{t_0^2}{\mathcal{A}_0\mathcal{A}_1}\frac{e^{-2\chi\left(\mathcal{K}z_{\perp}\right)^{\gamma}}}{1+\left[\frac{2v\Delta k}{\Omega-\lambda+\sqrt{4v^2\Delta k^2+(\Omega-\lambda)^2}}\right]^2},\\
			\left|T(2,A,-\vex{K}_1;1,B,\vex{K}_0+\vex{q})\right|^2\bigg|_{\vex{q}=-\vex{K}_1-2\vex{G}^{(1)}_2-[\vex{K}_0+\vex{G}^{(0)}_1-2\vex{G}^{(0)}_2]}\approx&\frac{t_0^2}{\mathcal{A}_0\mathcal{A}_1}\frac{e^{-2\chi\left(\mathcal{K}z_{\perp}\right)^{\gamma}}}{1+\left[\frac{2v\Delta k}{\Omega+\lambda+\sqrt{4v^2\Delta k^2+(\Omega+\lambda)^2}}\right]^2},
		\end{align}
	\end{subequations}
	where $|\vex{K}_1-\vex{K}_0|=0.437$\AA$^{-1}$, $\Delta k=|-\vex{K}_1-2\vex{G}^{(1)}_2-\vex{K}_0-\vex{G}^{(0)}_1+2\vex{G}^{(0)}_2|=0.416$\AA$^{-1}$ and $\mathcal{K}=|-\vex{K}_1-2\vex{G}^{(1)}_2|=4.51$\AA$^{-1}$. 
	
	\subsubsection{$\theta=5^{\circ}$}
	
	For $\theta=5^{\circ}$, there are only two distinct momenta. The spin-dependent interlayer tunneling terms are as follows:
	\begin{subequations}
		\begin{align}
			\left|T(2,A,\vex{K}_1;2,B,\vex{K}_0+\vex{q})\right|^2\bigg|_{\vex{q}=\vex{K}_1-\vex{K}_0}\approx& \frac{t_0^2}{\mathcal{A}_0\mathcal{A}_1}\frac{e^{-2\chi(|\vex{K}_1|z_{\perp})^{\gamma}}}{1+\left[\frac{2v|\vex{K}_1-\vex{K}_0|}{\Omega-\lambda+\sqrt{4v^2|\vex{K}_1-\vex{K}_0|^2+(\Omega-\lambda)^2}}\right]^2},\\
			\left|T(2,A,\vex{K}_1;1,B,\vex{K}_0+\vex{q})\right|^2\bigg|_{\vex{q}=\vex{K}_1-\vex{K}_0}\approx&\frac{t_0^2}{\mathcal{A}_0\mathcal{A}_1}\frac{e^{-2\chi(|\vex{K}_1|z_{\perp})^{\gamma}}}{1+\left[\frac{2v|\vex{K}_1-\vex{K}_0|}{\Omega+\lambda+\sqrt{4v^2|\vex{K}_1-\vex{K}_0|^2+(\Omega+\lambda)^2}}\right]^2},\\
			\left|T(2,A,-\vex{K}_1;2,B,\vex{K}_0+\vex{q})\right|^2\bigg|_{\vex{q}=-\vex{K}_1-2\vex{G}^{(1)}_2-[\vex{K}_0+\vex{G}^{(0)}_1-2\vex{G}^{(0)}_2]}\approx&
			\frac{t_0^2}{\mathcal{A}_0\mathcal{A}_1}\frac{e^{-2\chi\left(\mathcal{K}z_{\perp}\right)^{\gamma}}}{1+\left[\frac{2v\Delta k}{\Omega-\lambda+\sqrt{4v^2\Delta k^2+(\Omega-\lambda)^2}}\right]^2},\\
			\left|T(2,A,-\vex{K}_1;1,B,\vex{K}_0+\vex{q})\right|^2\bigg|_{\vex{q}=-\vex{K}_1-2\vex{G}^{(1)}_2-[\vex{K}_0+\vex{G}^{(0)}_1-2\vex{G}^{(0)}_2]}\approx&\frac{t_0^2}{\mathcal{A}_0\mathcal{A}_1}\frac{e^{-2\chi\left(\mathcal{K}z_{\perp}\right)^{\gamma}}}{1+\left[\frac{2v\Delta k}{\Omega+\lambda+\sqrt{4v^2\Delta k^2+(\Omega+\lambda)^2}}\right]^2},
		\end{align}
	\end{subequations}
	where $|\vex{K}_1-\vex{K}_0|=0.456$\AA$^{-1}$, $\Delta k=|-\vex{K}_1-2\vex{G}^{(1)}_2-\vex{K}_0-\vex{G}^{(0)}_1+2\vex{G}^{(0)}_2|=0.06$\AA$^{-1}$ and $\mathcal{K}=|-\vex{K}_1-2\vex{G}^{(1)}_2|=4.51$\AA$^{-1}$. Since the intervalley processes have a very small $\Delta k$, the intervalley contribution can be large. This explains the weak nonmonotonicity in ISOC an effective attraction for $\gamma=1$ and $\chi=0.2$.

	\subsubsection{$\theta=10^{\circ}$}
	
	For $\theta=10^{\circ}$, there are three distinct momenta. The spin-dependent interlayer tunneling terms are as follows:
	\begin{subequations}
		\begin{align}
			\left|T(2,A,\vex{K}_1;2,B,\vex{K}_0+\vex{q})\right|^2\bigg|_{\vex{q}=\vex{K}_1-\vex{K}_0}\approx& \frac{t_0^2}{\mathcal{A}_0\mathcal{A}_1}\frac{e^{-2\chi(|\vex{K}_1|z_{\perp})^{\gamma}}}{1+\left[\frac{2v|\vex{K}_1-\vex{K}_0|}{\Omega-\lambda+\sqrt{4v^2|\vex{K}_1-\vex{K}_0|^2+(\Omega-\lambda)^2}}\right]^2},\\
			\left|T(2,A,\vex{K}_1;1,B,\vex{K}_0+\vex{q})\right|^2\bigg|_{\vex{q}=\vex{K}_1-\vex{K}_0}\approx&\frac{t_0^2}{\mathcal{A}_0\mathcal{A}_1}\frac{e^{-2\chi(|\vex{K}_1|z_{\perp})^{\gamma}}}{1+\left[\frac{2v|\vex{K}_1-\vex{K}_0|}{\Omega+\lambda+\sqrt{4v^2|\vex{K}_1-\vex{K}_0|^2+(\Omega+\lambda)^2}}\right]^2},\\
			\left|T(2,A,-\vex{K}_1;2,B,\vex{K}_0+\vex{q})\right|^2\bigg|_{\vex{q}=-\vex{K}_1-\vex{G}^{(1)}_1-\vex{G}^{(1)}_2-[\vex{K}_0-\vex{G}^{(0)}_2]}\approx&
			\frac{t_0^2}{\mathcal{A}_0\mathcal{A}_1}\frac{e^{-2\chi\left(\mathcal{K}z_{\perp}\right)^{\gamma}}}{1+\left[\frac{2v\Delta k}{\Omega-\lambda+\sqrt{4v^2\Delta k^2+(\Omega-\lambda)^2}}\right]^2},\\
			\left|T(2,A,-\vex{K}_1;1,B,\vex{K}_0+\vex{q})\right|^2\bigg|_{\vex{q}=-\vex{K}_1-\vex{G}^{(1)}_1-\vex{G}^{(1)}_2-[\vex{K}_0-\vex{G}^{(0)}_2]}\approx&\frac{t_0^2}{\mathcal{A}_0\mathcal{A}_1}\frac{e^{-2\chi\left(\mathcal{K}z_{\perp}\right)^{\gamma}}}{1+\left[\frac{2v\Delta k}{\Omega+\lambda+\sqrt{4v^2\Delta k^2+(\Omega+\lambda)^2}}\right]^2},\\
			\left|T(2,A,-\vex{K}_1;2,B,\vex{K}_0+\vex{q})\right|^2\bigg|_{\vex{q}=-\vex{K}_1-2\vex{G}^{(1)}_2-[\vex{K}_0+\vex{G}^{(0)}_1-2\vex{G}^{(0)}_2]}\approx&
			\frac{t_0^2}{\mathcal{A}_0\mathcal{A}_1}\frac{e^{-2\chi\left(\mathcal{K}'z_{\perp}\right)^{\gamma}}}{1+\left[\frac{2v\Delta k'}{\Omega-\lambda+\sqrt{4v^2\Delta k'^2+(\Omega-\lambda)^2}}\right]^2},\\
			\left|T(2,A,-\vex{K}_1;1,B,\vex{K}_0+\vex{q})\right|^2\bigg|_{\vex{q}=-\vex{K}_1-2\vex{G}^{(1)}_2-[\vex{K}_0+\vex{G}^{(0)}_1-2\vex{G}^{(0)}_2]}\approx&\frac{t_0^2}{\mathcal{A}_0\mathcal{A}_1}\frac{e^{-2\chi\left(\mathcal{K}'z_{\perp}\right)^{\gamma}}}{1+\left[\frac{2v\Delta k'}{\Omega+\lambda+\sqrt{4v^2\Delta k'^2+(\Omega+\lambda)^2}}\right]^2},
		\end{align}
	\end{subequations}
	where $|\vex{K}_1-\vex{K}_0|=0.507$\AA$^{-1}$, $\Delta k=|-\vex{K}_1-\vex{G}^{(1)}_1-\vex{G}^{(1)}_2-\vex{K}_0+\vex{G}^{(0)}_2|=0.539$\AA$^{-1}$, $\mathcal{K}=|-\vex{K}_1-\vex{G}^{(1)}_1-\vex{G}^{(1)}_2|=3.41$\AA$^{-1}$, 
	$\Delta k'=|-\vex{K}_1-2\vex{G}^{(1)}_2-\vex{K}_0-\vex{G}^{(0)}_1+2\vex{G}^{(0)}_2|=0.383$\AA$^{-1}$, and $\mathcal{K}'=|-\vex{K}_1-2\vex{G}^{(1)}_2|=4.51$\AA$^{-1}$.

	\subsubsection{$\theta=15^{\circ}$}
	
	For $\theta=15^{\circ}$, there are two distinct momenta. The spin-dependent interlayer tunneling terms are as follows:
	\begin{subequations}
		\begin{align}
			\left|T(2,A,\vex{K}_1;2,B,\vex{K}_0+\vex{q})\right|^2\bigg|_{\vex{q}=\vex{K}_1-\vex{K}_0}\approx& \frac{t_0^2}{\mathcal{A}_0\mathcal{A}_1}\frac{e^{-2\chi(|\vex{K}_1|z_{\perp})^{\gamma}}}{1+\left[\frac{2v|\vex{K}_1-\vex{K}_0|}{\Omega-\lambda+\sqrt{4v^2|\vex{K}_1-\vex{K}_0|^2+(\Omega-\lambda)^2}}\right]^2},\\
			\left|T(2,A,\vex{K}_1;1,B,\vex{K}_0+\vex{q})\right|^2\bigg|_{\vex{q}=\vex{K}_1-\vex{K}_0}\approx&\frac{t_0^2}{\mathcal{A}_0\mathcal{A}_1}\frac{e^{-2\chi(|\vex{K}_1|z_{\perp})^{\gamma}}}{1+\left[\frac{2v|\vex{K}_1-\vex{K}_0|}{\Omega+\lambda+\sqrt{4v^2|\vex{K}_1-\vex{K}_0|^2+(\Omega+\lambda)^2}}\right]^2},\\
			\left|T(2,A,-\vex{K}_1;2,B,\vex{K}_0+\vex{q})\right|^2\bigg|_{\vex{q}=-\vex{K}_1-\vex{G}^{(1)}_1-\vex{G}^{(1)}_2-[\vex{K}_0-\vex{G}^{(0)}_2]}\approx&
			\frac{t_0^2}{\mathcal{A}_0\mathcal{A}_1}\frac{e^{-2\chi\left(\mathcal{K}z_{\perp}\right)^{\gamma}}}{1+\left[\frac{2v\Delta k}{\Omega-\lambda+\sqrt{4v^2\Delta k^2+(\Omega-\lambda)^2}}\right]^2},\\
			\left|T(2,A,-\vex{K}_1;1,B,\vex{K}_0+\vex{q})\right|^2\bigg|_{\vex{q}=-\vex{K}_1-\vex{G}^{(1)}_1-\vex{G}^{(1)}_2-[\vex{K}_0-\vex{G}^{(0)}_2]}\approx&\frac{t_0^2}{\mathcal{A}_0\mathcal{A}_1}\frac{e^{-2\chi\left(\mathcal{K}z_{\perp}\right)^{\gamma}}}{1+\left[\frac{2v\Delta k}{\Omega+\lambda+\sqrt{4v^2\Delta k^2+(\Omega+\lambda)^2}}\right]^2},
		\end{align}
	\end{subequations}
	where $|\vex{K}_1-\vex{K}_0|=0.581$\AA$^{-1}$, $\Delta k=|-\vex{K}_1-\vex{G}^{(1)}_1-\vex{G}^{(1)}_2-\vex{K}_0+\vex{G}^{(0)}_2|=0.249$\AA$^{-1}$, and $\mathcal{K}=|-\vex{K}_1-\vex{G}^{(1)}_1-\vex{G}^{(1)}_2|=3.41$\AA$^{-1}$.

	\subsubsection{$\theta=20^{\circ}$}
	
	For $\theta=20^{\circ}$, there are three distinct momenta. The spin-dependent interlayer tunneling terms are as follows:
	\begin{subequations}
		\begin{align}
			\left|T(2,A,-\vex{K}_1;2,B,\vex{K}_0+\vex{q})\right|^2\bigg|_{\vex{q}=-\vex{K}_1-\vex{G}^{(1)}_1-\vex{G}^{(1)}_2-[\vex{K}_0-\vex{G}^{(0)}_2]}\approx&
			\frac{t_0^2}{\mathcal{A}_0\mathcal{A}_1}\frac{e^{-2\chi\left(\mathcal{K}z_{\perp}\right)^{\gamma}}}{1+\left[\frac{2v\Delta k}{\Omega-\lambda+\sqrt{4v^2\Delta k^2+(\Omega-\lambda)^2}}\right]^2},\\
			\left|T(2,A,-\vex{K}_1;1,B,\vex{K}_0+\vex{q})\right|^2\bigg|_{\vex{q}=-\vex{K}_1-\vex{G}^{(1)}_1-\vex{G}^{(1)}_2-[\vex{K}_0-\vex{G}^{(0)}_2]}\approx&\frac{t_0^2}{\mathcal{A}_0\mathcal{A}_1}\frac{e^{-2\chi\left(\mathcal{K}z_{\perp}\right)^{\gamma}}}{1+\left[\frac{2v\Delta k}{\Omega+\lambda+\sqrt{4v^2\Delta k^2+(\Omega+\lambda)^2}}\right]^2},\\
			\left|T(2,A,\vex{K}_1;2,B,\vex{K}_0+\vex{q})\right|^2\bigg|_{\vex{q}=\vex{K}_1-\vex{G}^{(1)}_1-[\vex{K}_0-\vex{G}^{(0)}_1-\vex{G}^{(0)}_2]}\approx&
			\frac{t_0^2}{\mathcal{A}_0\mathcal{A}_1}\frac{e^{-2\chi\left(\mathcal{K}'z_{\perp}\right)^{\gamma}}}{1+\left[\frac{2v\Delta k'}{\Omega-\lambda+\sqrt{4v^2\Delta k'^2+(\Omega-\lambda)^2}}\right]^2},\\
			\left|T(2,A,\vex{K}_1;2,B,\vex{K}_0+\vex{q})\right|^2\bigg|_{\vex{q}=\vex{K}_1-\vex{G}^{(1)}_1-[\vex{K}_0-\vex{G}^{(0)}_1-\vex{G}^{(0)}_2]}\approx&
			\frac{t_0^2}{\mathcal{A}_0\mathcal{A}_1}\frac{e^{-2\chi\left(\mathcal{K}'z_{\perp}\right)^{\gamma}}}{1+\left[\frac{2v\Delta k'}{\Omega+\lambda+\sqrt{4v^2\Delta k'^2+(\Omega+\lambda)^2}}\right]^2},\\
			\left|T(2,A,\vex{K}_1;2,B,\vex{K}_0+\vex{q})\right|^2\bigg|_{\vex{q}=\vex{K}_1-\vex{G}^{(1)}_2-[\vex{K}_0+\vex{G}^{(0)}_1-2\vex{G}^{(0)}_2]}\approx&
			\frac{t_0^2}{\mathcal{A}_0\mathcal{A}_1}\frac{e^{-2\chi\left(\mathcal{K}''z_{\perp}\right)^{\gamma}}}{1+\left[\frac{2v\Delta k''}{\Omega-\lambda+\sqrt{4v^2\Delta k''^2+(\Omega-\lambda)^2}}\right]^2},\\
			\left|T(2,A,\vex{K}_1;2,B,\vex{K}_0+\vex{q})\right|^2\bigg|_{\vex{q}=\vex{K}_1-\vex{G}^{(1)}_2-[\vex{K}_0+\vex{G}^{(0)}_1-2\vex{G}^{(0)}_2]}\approx&
			\frac{t_0^2}{\mathcal{A}_0\mathcal{A}_1}\frac{e^{-2\chi\left(\mathcal{K}''z_{\perp}\right)^{\gamma}}}{1+\left[\frac{2v\Delta k''}{\Omega+\lambda+\sqrt{4v^2\Delta k''^2+(\Omega+\lambda)^2}}\right]^2},
		\end{align}
	\end{subequations}
	where $\Delta k=|-\vex{K}_1-\vex{G}^{(1)}_1-\vex{G}^{(1)}_2-\vex{K}_0+\vex{G}^{(0)}_2|=0.0778$\AA$^{-1}$, $\mathcal{K}=|-\vex{K}_1-\vex{G}^{(1)}_1-\vex{G}^{(1)}_2|=3.41$\AA$^{-1}$, 
	$\Delta k'=|\vex{K}_1-\vex{G}^{(1)}_1-\vex{K}_0+\vex{G}^{(0)}_1+\vex{G}^{(0)}_2|=0.562$\AA$^{-1}$, $\mathcal{K}'=|\vex{K}_1-\vex{G}^{(1)}_1|=4.51$\AA$^{-1}$, 
	$\Delta k''=|\vex{K}_1-\vex{G}^{(1)}_2-\vex{K}_0-\vex{G}^{(0)}_1+2\vex{G}^{(0)}_2|=0.556$\AA$^{-1}$, $\mathcal{K}''=|\vex{K}_1-\vex{G}^{(1)}_2|=4.51$\AA$^{-1}$. At $\theta=20^{\circ}$, the intervalley processes have smaller crystal momentum than the intravalley processes. In addition, $\Delta k$ for the intervalley tunneling is quite small. This is why we find $\lambda_I<0$ at $\theta=20^{\circ}$. Another important point is that the tunneling crystal momentum is no longer at the first Brillouin zone corner (i.e., $K$ or $-K$). This explains the small effective attraction for $\theta\ge 20^{\circ}$
	
	\subsubsection{$\theta=25^{\circ}$}
	
	For $\theta=25^{\circ}$, there are two distinct momenta. The spin-dependent interlayer tunneling terms are as follows:
	\begin{subequations}
		\begin{align}
			\left|T(2,A,-\vex{K}_1;2,B,\vex{K}_0+\vex{q})\right|^2\bigg|_{\vex{q}=-\vex{K}_1-\vex{G}^{(1)}_1-\vex{G}^{(1)}_2-[\vex{K}_0-\vex{G}^{(0)}_2]}\approx&
			\frac{t_0^2}{\mathcal{A}_0\mathcal{A}_1}\frac{e^{-2\chi\left(\mathcal{K}z_{\perp}\right)^{\gamma}}}{1+\left[\frac{2v\Delta k}{\Omega-\lambda+\sqrt{4v^2\Delta k^2+(\Omega-\lambda)^2}}\right]^2},\\
			\left|T(2,A,-\vex{K}_1;1,B,\vex{K}_0+\vex{q})\right|^2\bigg|_{\vex{q}=-\vex{K}_1-\vex{G}^{(1)}_1-\vex{G}^{(1)}_2-[\vex{K}_0-\vex{G}^{(0)}_2]}\approx&\frac{t_0^2}{\mathcal{A}_0\mathcal{A}_1}\frac{e^{-2\chi\left(\mathcal{K}z_{\perp}\right)^{\gamma}}}{1+\left[\frac{2v\Delta k}{\Omega+\lambda+\sqrt{4v^2\Delta k^2+(\Omega+\lambda)^2}}\right]^2},\\
			\left|T(2,A,\vex{K}_1;2,B,\vex{K}_0+\vex{q})\right|^2\bigg|_{\vex{q}=\vex{K}_1-\vex{G}^{(1)}_2-[\vex{K}_0+\vex{G}^{(0)}_1-2\vex{G}^{(0)}_2]}\approx&
			\frac{t_0^2}{\mathcal{A}_0\mathcal{A}_1}\frac{e^{-2\chi\left(\mathcal{K}''z_{\perp}\right)^{\gamma}}}{1+\left[\frac{2v\Delta k''}{\Omega-\lambda+\sqrt{4v^2\Delta k''^2+(\Omega-\lambda)^2}}\right]^2},\\
			\left|T(2,A,\vex{K}_1;2,B,\vex{K}_0+\vex{q})\right|^2\bigg|_{\vex{q}=\vex{K}_1-\vex{G}^{(1)}_2-[\vex{K}_0+\vex{G}^{(0)}_1-2\vex{G}^{(0)}_2]}\approx&
			\frac{t_0^2}{\mathcal{A}_0\mathcal{A}_1}\frac{e^{-2\chi\left(\mathcal{K}''z_{\perp}\right)^{\gamma}}}{1+\left[\frac{2v\Delta k''}{\Omega+\lambda+\sqrt{4v^2\Delta k''^2+(\Omega+\lambda)^2}}\right]^2},
		\end{align}
	\end{subequations}
	where $\Delta k=|-\vex{K}_1-\vex{G}^{(1)}_1-\vex{G}^{(1)}_2-\vex{K}_0+\vex{G}^{(0)}_2=0.352|$\AA$^{-1}$, $\mathcal{K}=|-\vex{K}_1-\vex{G}^{(1)}_1-\vex{G}^{(1)}_2|=3.41$\AA$^{-1}$, 
	$\Delta k''=|\vex{K}_1-\vex{G}^{(1)}_2-\vex{K}_0-\vex{G}^{(0)}_1+2\vex{G}^{(0)}_2|=0.168$\AA$^{-1}$, $\mathcal{K}''=|\vex{K}_1-\vex{G}^{(1)}_2|=4.51$\AA$^{-1}$.	Similar to $\theta=20^{\circ}$, the intervalley contributions are stronger than intravalley contributions, and $\lambda_I<0$ (albeit vanishingly small in some cases).
	
	\subsubsection{$\theta=30^{\circ}$}
	
	For $\theta=30^{\circ}$, there is actually only one distinct momentum, but we check both intravalley and intervalley processes. The spin-dependent interlayer tunneling terms are as follows:
	\begin{subequations}
		\begin{align}
			\left|T(2,A,-\vex{K}_1;2,B,\vex{K}_0+\vex{q})\right|^2\bigg|_{\vex{q}=-\vex{K}_1-2\vex{G}^{(1)}_2-[\vex{K}_0+2\vex{G}^{(0)}_1-2\vex{G}^{(0)}_2]}\approx&
			\frac{t_0^2}{\mathcal{A}_0\mathcal{A}_1}\frac{e^{-2\chi\left(\mathcal{K}z_{\perp}\right)^{\gamma}}}{1+\left[\frac{2v\Delta k}{\Omega-\lambda+\sqrt{4v^2\Delta k^2+(\Omega-\lambda)^2}}\right]^2},\\
			\left|T(2,A,-\vex{K}_1;1,B,\vex{K}_0+\vex{q})\right|^2\bigg|_{\vex{q}=-\vex{K}_1-2\vex{G}^{(1)}_2-[\vex{K}_0+2\vex{G}^{(0)}_1-2\vex{G}^{(0)}_2]}\approx&\frac{t_0^2}{\mathcal{A}_0\mathcal{A}_1}\frac{e^{-2\chi\left(\mathcal{K}z_{\perp}\right)^{\gamma}}}{1+\left[\frac{2v\Delta k}{\Omega+\lambda+\sqrt{4v^2\Delta k^2+(\Omega+\lambda)^2}}\right]^2},\\
			\left|T(2,A,\vex{K}_1;2,B,\vex{K}_0+\vex{q})\right|^2\bigg|_{\vex{q}=\vex{K}_1-\vex{G}^{(1)}_2-[\vex{K}_0+\vex{G}^{(0)}_1-2\vex{G}^{(0)}_2]}\approx&
			\frac{t_0^2}{\mathcal{A}_0\mathcal{A}_1}\frac{e^{-2\chi\left(\mathcal{K}''z_{\perp}\right)^{\gamma}}}{1+\left[\frac{2v\Delta k''}{\Omega-\lambda+\sqrt{4v^2\Delta k''^2+(\Omega-\lambda)^2}}\right]^2},\\
			\left|T(2,A,\vex{K}_1;2,B,\vex{K}_0+\vex{q})\right|^2\bigg|_{\vex{q}=\vex{K}_1-\vex{G}^{(1)}_2-[\vex{K}_0+\vex{G}^{(0)}_1-2\vex{G}^{(0)}_2]}\approx&
			\frac{t_0^2}{\mathcal{A}_0\mathcal{A}_1}\frac{e^{-2\chi\left(\mathcal{K}''z_{\perp}\right)^{\gamma}}}{1+\left[\frac{2v\Delta k''}{\Omega+\lambda+\sqrt{4v^2\Delta k''^2+(\Omega+\lambda)^2}}\right]^2},
		\end{align}
	\end{subequations}
	where $\Delta k=|\vex{q}=-\vex{K}_1-2\vex{G}^{(1)}_2-\vex{K}_0-2\vex{G}^{(0)}_1+2\vex{G}^{(0)}_2|=0.245$\AA$^{-1}$, $\mathcal{K}=|-\vex{K}_1-2\vex{G}^{(1)}_2|=4.51$\AA$^{-1}$, 
	$\Delta k''=|\vex{K}_1-\vex{G}^{(1)}_2-\vex{K}_0-\vex{G}^{(0)}_1+2\vex{G}^{(0)}_2|=0.245$\AA$^{-1}$, and $\mathcal{K}''=|\vex{K}_1-\vex{G}^{(1)}_2|=4.51$\AA$^{-1}$.	$\lambda_I=0$ at $\theta=30^{\circ}$.

	\begin{figure}[t!]
		\includegraphics[width=0.8\textwidth]{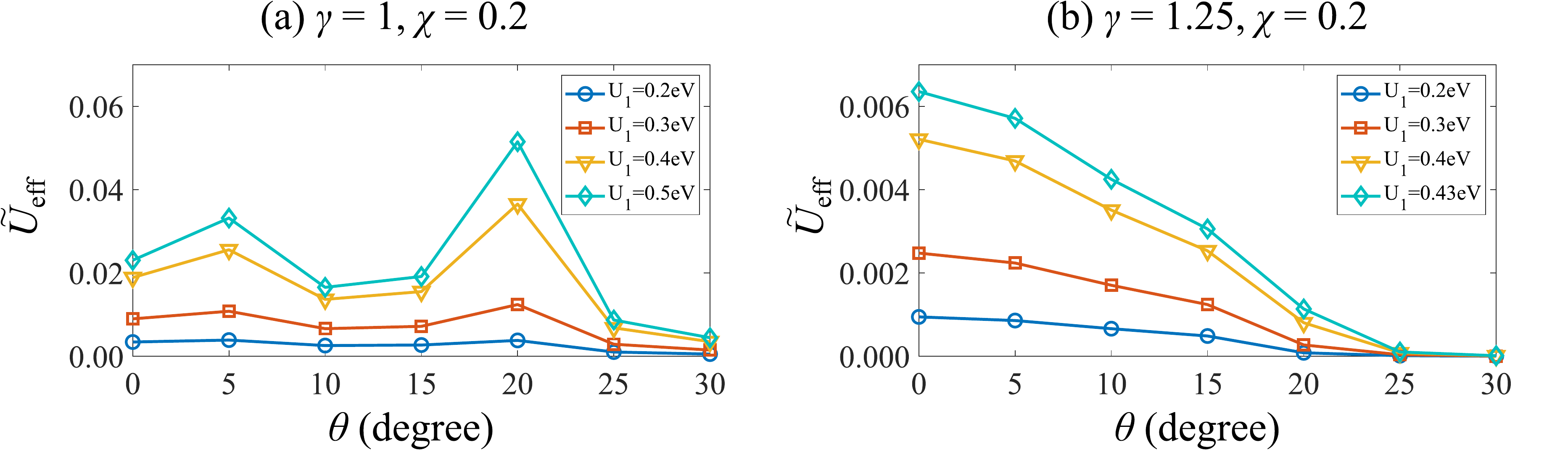}
		\caption{Effective attraction for different values of $U_1$. We plot the dimensionless spin-orbit coupling $\tilde{U}_{\text{eff}}=\mathcal{U}_{\text{eff}} \mathcal{A}_0\mathcal{A}_1 \text{eV}/t_0^2$ as a function of $\theta$ with different values of $U_1$. We use $\gamma=1.25$ and $\chi=0.2$ in all the curves.}
		\label{Fig:Att_vs_U1}
	\end{figure}	
	
	\subsection{Effective attraction versus $U_1$}
	
	Here, we discuss $U_1$ (short-range Coulomb interaction) dependence in the effective attraction strength $\mathcal{U}_{\text{eff}}$. In Fig.~\ref{Fig:Att_vs_U1}, we plot $\tilde{U}_{\text{eff}}=\mathcal{U}_{\text{eff}} \mathcal{A}_0\mathcal{A}_1 \text{eV}/t_0^2$ (the dimensionless attraction) as a function of $\theta$ with (a) $\gamma=1$, $\chi=0.2$ and (b) $\gamma=1.25$, $\chi=0.2$ , and different values of $U_1$ ranging from 0.2eV to 0.43eV. ($U_1\le E_B$ in our theory, where $E_B\approx 0.43$eV.) The qualitative trends are very similar because the values of $U_1$ do not generate extremely small denominators in second order perturbation theory. However, the results show quantitative dependence in $U_1$, and a larger $U_1$ generally enhances $\tilde{U}_{\text{eff}}$.


\begin{thebibliography}{54}
\bibitem[{zpe()}]{zperp}
		\bibinfo{note}{For simplicity, we assume $z_{\perp}=3.34${\AA }, the same as
			the distance between two graphene layers. The interlayer distance dependence
			can be absorbed into the parameter $\chi$.}
\bibitem[{par()}]{parameters}
		\bibinfo{note}{In Ref.~\cite{Bistritzer2010}, $\gamma=1.25$ and $\chi=0.13$
			were used for twisted bilayer graphene. We assume that $\gamma$ and $\chi$
			here are not significantly different from those values.}
\bibitem[{pai()}]{pairing_symmetry}
		\bibinfo{note}{The $\mathcal{U}_{\text{eff}}$ interaction [Eq.~(7)] preserves
			$SU(2)\times SU(2)$ symmetry, but this is an artifact of our minimal model.
			Usually, the short-range interaction preserves only the global $SU(2)$
			symmetry, and the higher-order contributions involving both $V_A$ and $V_B$
			tunnelings break the spin symmetry entirely.}
\end{thebibliography}

\end{document}